\documentclass{article}
\usepackage[a4paper, total={6in, 8in}]{geometry}

\usepackage{amsmath}
\usepackage{xcolor}
\usepackage{lastpage}
\usepackage{fancyhdr}
\usepackage{graphicx}
\usepackage{url}
\usepackage{hyperref}
\usepackage{myStyle}
\usepackage{caption}
\usepackage{subcaption}
\usepackage{mathtools}
\usepackage[title]{appendix}

\newcommand{\viz}{\textit{viz. }}
\newcommand{\credits}[1]{(\textit{Adopted from:#1})}

\usepackage{lineno}

\begin{document}

\title{Pulse Shape Simulation and Discrimination using Machine Learning Techniques}

\author{S. Dutta\authormark{1*}, S. Ghosh\authormark{3}, S. Bhattacharya\authormark{1}, and S. Saha\authormark{2}}

\address{\authormark{1}High Energy Nuclear and Particle Physics Division, Saha Institute of Nuclear Physics - a CI of Homi Bhabha National Institute, Kolkata - 700064, INDIA\\
\authormark{2}Applied Nuclear Physics Division, 
Saha Institute of Nuclear Physics - a CI of Homi Bhabha National Institute, Kolkata - 700064, INDIA\\
\authormark{3}Department of Physics and Astronomy, Purdue University, West Lafayette, IN, 47907, USA}

\email{\authormark{*}shubhamdutta$\_$16@yahoo.com} 

\begin{abstract}
\noindent
An essential metric for the quality of a particle-identification experiment is its statistical power to discriminate between signal and background. Pulse shape discrimination (PSD) is a basic method for this purpose in many nuclear, high-energy and rare-event search experiments where scintillation detectors are used. Conventional techniques exploit the difference between decay-times of the pulses from signal and background events or pulse signals caused by different types of radiation quanta to achieve good discrimination. However, such techniques are efficient only when the total light-emission is sufficient to get a proper pulse profile. This is only possible when adequate amount of energy is deposited from recoil of the electrons or the nuclei of the scintillator materials caused by the incident particle on the detector. But, rare-event search experiments like direct search for dark matter do not always satisfy these conditions. Hence, it becomes imperative to have a method that can deliver a very efficient discrimination in these scenarios. Neural network based machine-learning algorithms have been used for classification problems in many areas of physics especially in high-energy experiments and have given better results compared to conventional techniques. We present the results of our investigations of two network based methods \viz Dense Neural Network and Recurrent Neural Network, for pulse shape discrimination and compare the same with conventional methods.
\end{abstract}

\section{Introduction}
\label{introduction} 
Pulse shape discrimination (PSD) have been widely used to discriminate between the different radiation quanta, such as photons (X-rays, gamma rays etc), electrons, neutrons, protons, alpha particles, etc. As these particles interact with the medium, they leave traces of signals (pulses) that differ in shape characterized by rise time, fall time, charge content and various other parameters. In a mixed radiation field experiment, PSD is considered as indispensable method to discriminate between the signal from the radiation quanta of interest and the background. One of the of first techniques for PSD was developed using the time domain information \cite{Storey1958}. Since then various PSD techniques have been evolved and utilized, mostly in the time domain, such as charge integration\cite{CImethod_Sabbah1968}, mean-time\cite{Lee2014}, zero cross-over\cite{ZeroX_Roush1964}, pulse gradient\cite{PGA_DMellow2007}, time-over-threshold\cite{ToT_Kipnis1997}\cite{ToT_Ngyren1991}, etc. In addition, PSD in the frequency domain has also been demonstrated by digital signal processing techniques such as discrete wavelet transform and found to perform better than conventional time domain technique for discrimination between neutrons and gamma-rays using liquid organic scintillator\cite{Yousefi2009}.\\

\noindent
Inorganic scintillators are quite often used in radiation detection for their much higher light output as compared to the organic scintillators. Manifestation of photo-peaks in inorganic scintillators helps in carrying out nuclear spectroscopic investigation in the energy domain. Thallium doped Cesium Iodide [CsI(Tl)] scintillator has been used across the energy domain for spectroscopic as well as calorimetric investigation exploiting the PSD techniques for improved particle identification. This method has been widely used for discrimination between light charged particles at $E \leq 20\,{\rm MeV}$ in nuclear reaction studies at low and intermediate energy domain \cite{CHIMERA2002} and proposed to be utilized as a new technique to discriminate between the hadronic and the electromagnetic showers in the high energy domain at the $e^+e^-$ collider experiment Belle II \cite{Longo2020}.\\

\noindent
When an energetic particle passes through a scintillator, the amount of light produced is dependent on the energy-loss per unit path-length ($dE/dx$) or specific energy loss (SEL) of the incident particle. This quantity varies with energy and the incident particle-type due to the difference in the nature of interaction of the incident particle with the scintillating medium. As the particle loses energy in the medium, the SEL changes resulting in variation in the intensity of light-output as a function of time. Hence, scintillators have the property of producing pulses with different decay-times based on the incident particle-type.  This is the underlying principle acting as the basis for pulse shape discrimination.\\

\noindent
Inorganic scintillators have also been used because of their high light output and variation of the pulse shapes by the interaction of different hadronic and electromagnetic particles in experiments searching for rare events, such as dark matter search (DMS) experiment using CsI(Tl) scintillator\cite{KIMS2002}, neutrinoless double beta decay experiment using $^{48}$Ca enriched Calcium Fluoride (${\rm CaF}_2$) scintillator\cite{Umehara2015}, etc. The DMS experiment, in particular, looks for a very small amount of energy (of the order of a few keV to ~100 keV) dumped into the scintillator by nuclear recoil suffered due to very rare interaction of the dark matter particles with its nuclei. A major challenge in the DMS experiment is the overwhelming presence of unwanted background caused by the electrons, muons, gamma rays, neutrons, etc. \cite{Ghosh2022} which must be filtered out. PSD has been effectively used\cite{KIMS2014} to filter out the unwanted electromagnetic background. Furthermore, the amount of kinetic energy imparted to the recoiling nucleus becomes less as the DMS experiments look for lighter DM candidates (sub-GeV to a few GeV of mass). Efficient discrimination by the conventional PSD techniques involving time or frequency domain information becomes increasingly constrained at such tiny recoil energy domain. In this work, it is demonstrated that the machine learning (ML) technique using effective algorithms has the potential of better discrimination than the conventional methods.\\

\noindent
\textcolor{blue}{ML} is a class of computer algorithms that ``train'' on existing data to learn its general features in order to be able to predict the outcome, when new data is fed to it. To be specific, this pertains to supervised learning, nonetheless ML is also capable of doing unsupervised learning. This is akin to fitting a dataset with a function by adjusting the function-parameters, and then using the fitted-function to predict new outcomes. However, ML is much more diverse in the sense that one need not provide any functional form, a priori. ML itself finds the best function that captures the general trend of the dataset while training. Here, we have explored PSD with Boosted Decision Trees (BDT) \cite{Freund1997} and network-based ML \cite{Goodfellow2016}.\\

\noindent
In this paper, we confine our attention to the task of discrimination of scintillation signals caused by gamma rays predominantly through the electron recoil, termed as electron scintillation (ES) and that due to nuclear recoil caused by the neutrons, which is termed as nuclear scintillation (NS) in a couple of inorganic crystalline scintillators such as CsI(Tl) and Bismuth Germanate (BGO).\\

\noindent
The paper is organized as follows. Section 2 introduces the nuances and relevant details of the ML techniques followed. GEANT4 simulation, as applied to the task of pulse shape discrimination, has been discussed in Section 3. Simulation deals with the scintillation caused by gamma rays predominantly through the electron recoil, termed as electron scintillation (ES), such as CsI(Tl) and Bismuth Germanate (BGO). Experimental details and training of data for ML have been described in Section 4. Analysis of results and comparison of ML based results with a few conventional PSD methods have been discussed in Section 5, followed by conclusion.

\section{Machine Learning Methods}
\label{MLbasedMethods}

\noindent
\textcolor{blue}ML based methods have shown dramatic improvements in classification problems \cite{ImageNet_CNN}. These algorithms have become ubiquitous through their wide range of applications, starting from facial-recognition to self-driving cars. In physics, they have been applied for particle identification problems in high energy physics \cite{CMS2018} \cite{CMS2020} \cite{DUNE2020} \cite{Aurisano2016} \cite{Soham2022} \cite{WZTagNN_Chen2020} and have given better results compared to conventional methods. In the following, we mention the important parameters of the network models used specifically for the current work.\\

\begin{figure}[ht!]
    \centering
	\includegraphics[width=140mm]{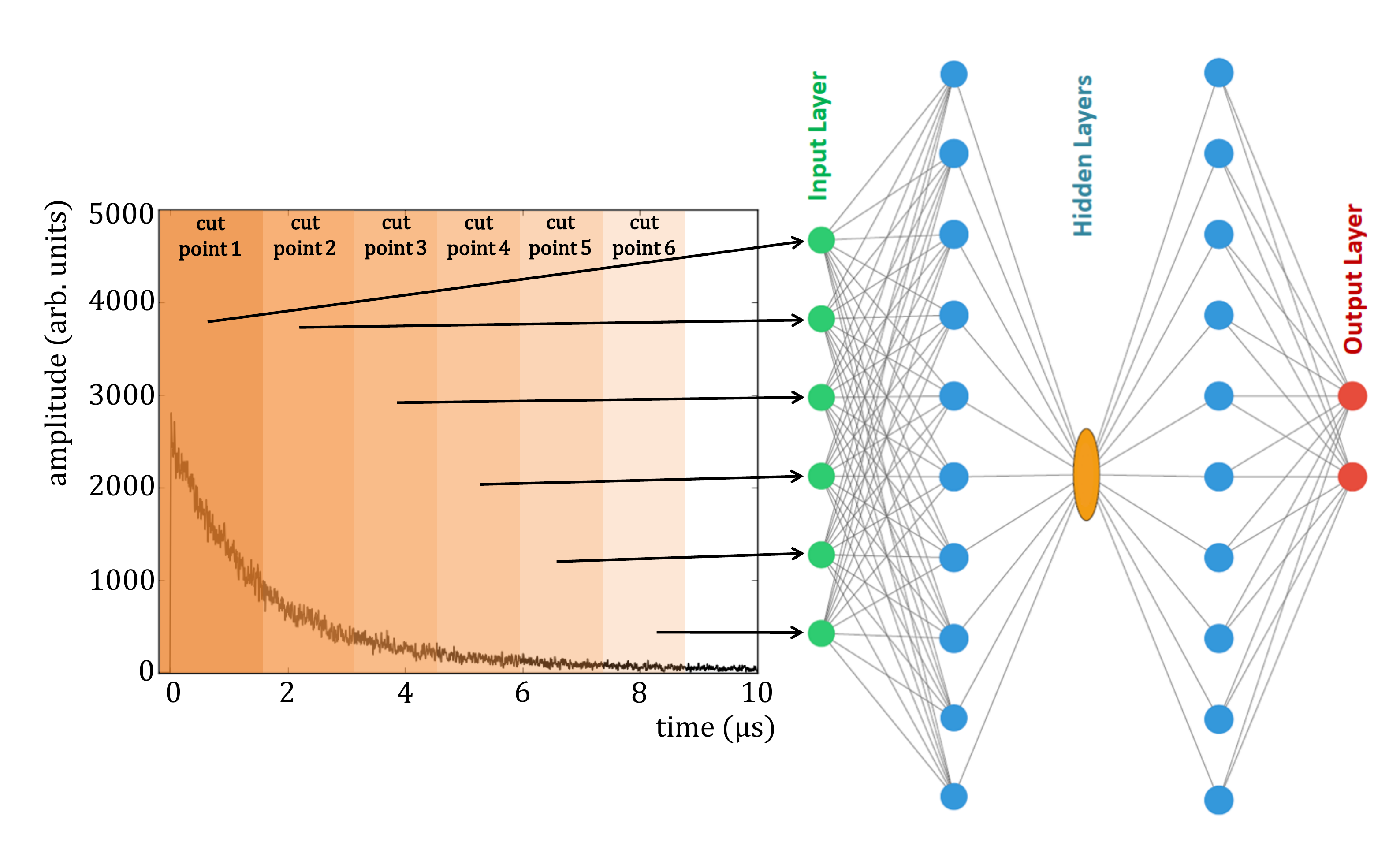}
	\caption{\label{PSDdenseNetwork} Dense Network for PSD}
\end{figure}

\noindent
A basic \textbf{Dense Neural Network (DNN)} was made to serve as a benchmark for the performance of  network-based methods. The network consisted of 6 input nodes, 2 hidden layers having 10 nodes each and an output layer having 2 nodes. It is called a dense network because all nodes of a particular layer are connected to all the nodes from the previous layer. The schematic of the network is shown in figure \ref{PSDdenseNetwork}. The total number of training parameters in the network is 202. Dropouts \cite{dropout_Srivastava2014} were added in order to prevent over-training of the network. The drop-out fraction is 0.2. The input to the network is the area under the pulse starting from t=0 to various cut-points in time as shown in figure \ref{PSDdenseNetwork}. Calculating the value for the input nodes in this way ensures that the input values to the network remains between $0-1$. This does away the need for scaling the input before feeding it to the network. All the nodes have rectified linear unit (ReLU) \cite{Nair2010} activations except for the nodes in the last (output) layer which have softmax activations \cite{Goodfellow2016}. The trained-network assigns a score in the range $0$\textendash$1$ to every pulse. A score of $0$ indicates that the network classifies the pulse to belong to background with $100\%$ certainty, while a score of $1$ means it is classified as signal.\\

\noindent
Next, we tried a \textbf{Recurrent Neural Network} (RNN) for the purpose, since the pulse-data that is being given as input to the network is a time-series data. Thus, each data-point in time is related with the previous data-point. The memory feature of the RNN can exploit this correlation and learn to distinguish the time-series dataset better. It has 5 LSTM units followed by a dense connected hidden layer with 15 nodes. The total number of training parameters is 262. The entire time-slice data is fed to each LSTM unit in the input in the form of a vector.

\section{Detector Simulation}
\label{detectorSimulation}

\noindent
The detector simulation to produce pulses serves broadly two purposes. First, a detailed simulation helps to mimic the actual detector response that can aid in designing, evaluating and optimizing detector shielding. For rare-event search experiments, detector shielding is a very important aspect as it helps to further cut-down the background events. This can then be corroborated by performing experiments at the site and comparing the results with simulation. Second, it enables us to have a large dataset to train and validate the performance of the machine learning (ML) algorithms. Simulation also gives us handles to simulate any scenario at will, which is essential for doing proper systematic studies. \\

\noindent
GEANT4 \cite{geant4} package is used to simulate the production and transportation of scintillation photons when $\gamma$-rays enter and deposit energy in the scintillators. The PMT-response is simulated using a separate code that takes the output from GEANT4 to produce the final scintillation pulse. Scintillation photons are generated in GEANT4 when the parameters that characterize the scintillation light-output are provided to it. The scintillation spectrum, rise-time, the fast and slow decay-time constants and light yield are the main parameters. The variation of the input distribution that is fed to the network as these parameters are varied is detailed in the Appendix \ref{App_VariationOfInput}. The spectrum is provided as a list of wavelengths and corresponding intensity relative to the maximum. There is provision to specify the fast and slow components of the spectrum corresponding to the fast and slow decay-times. GEANT4 generates scintillation photons based on these parameters having isotropic distribution and linear polarization. However, this is only the generation part of the photons.\\ 

\begin{figure}[ht!]
    \centering
	\includegraphics[width=50mm]{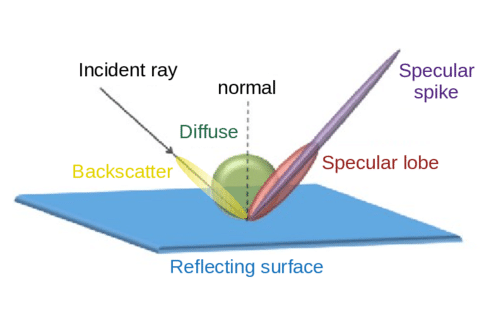}
	\caption{\label{unifiedModel} Unified model of reflection in GEANT4  \credits{\cite{unifiedModel_ref}}}
\end{figure}

\noindent
The simulation of photon transportation is carried out when the refractive index of the medium is provided. In order to simulate processes at the interface of two media, the surface boundary needs to be defined. This definition includes the surface roughness and the nature of the boundary as dielectric-dielectric or dielectric-metal. This is particularly important for reflection of photons at the boundary surface. Under the unified model \cite{unifiedModel_ref}, reflectance has 4 main types \viz specular spike, specular lobe, backscatter and Lambertian (diffused) (figure \ref{unifiedModel}). For each wavelength that was given in the scintillation spectrum, a list of 4 numbers is provided. These numbers represent the probability for a photon, of a particular wavelength, to be reflected following one of the 4 types. In case the boundary is in between two transparent media, then the refractive index is used to calculate the total internal reflection.\\

\noindent
The reflectance of teflon is modeled as having 99\% reflectivity \cite{Silva2010}, remaining 1\% is transmitted or absorbed. Among the photons that are reflected, the main type of reflection is diffused (90\%) and the rest is specular lobe (10\%). The detection of a photon is simulated by changing the status of a flag (associated with the photon) to `DETECTED' or `ABSORBED'. This is implemented both as boundary process and bulk process within the PMT glass window. GEANT4 provides a handle, termed as `EFFICIENCY', which is a number corresponding to each wavelength that quantifies the detection efficiency (quantum efficiency) of the PMT. \\

\noindent
The PMT simulation was done using a separate private code that implements the PMT response. There are mainly two components at play here. First is the gain of the PMT and second is the transit time spread (TTS). However, for a slow scintillator (compared to TTS of PMT) the TTS does not play a big role. The gain corresponding to a photon is sampled from a distribution as described in \cite{pmtGain_Rademacker2002}. The final pulse is produced by convoluting the scintillation pulse obtained from GEANT4 with the PMT response that includes the gain and TTS. If a profile of the electrical noise is available for the setup that too can be included here.

\section{Experimental and Training Data}
\label{experimentalAndTrainingData}

\noindent
The experimental data is obtained from two detector setups. In the present work this was done for BGO and CsI(Tl) scintillators. Experimental validation of the simulation of both setups gives us confidence that the simulation can be tuned for different detectors. In both the cases, Cesium-137 ($^{137}$Cs) source is used to obtain the deposited-energy spectrum. $^{137}$Cs is a radioactive source which emits mono-energetic $\gamma$-rays of 662 keV energy. Mono-energetic $\gamma$ ray was used in order to fix the input energy parameter of the radiation quanta (within the limits of energy resolution of the detector), so that we can use the pulse shape parameters (rise time and decay time) as variables for the study.\\

\noindent
The BGO data was taken from \cite{Ogawara2016} for the first setup. The $^{137}$Cs $\gamma$-source is placed in front of a cylindrical BGO crystal having 2 mm thickness $\times$ 13 mm diameter. The energy spectrum is obtained using integrated charge of pulse that is acquired using a digital oscilloscope (LeCroy WaveRunner 64xi).\\ 

\noindent
In the second case, a cylindrical CsI(Tl) crystal having 50 mm thickness $\times$ 50 mm diameter is studied. The Cs$^{137}$ source is used to obtain the energy-spectrum and pulse-data, as mentioned earlier, with which the simulated data is compared. The two end-faces of the crystal were optically coupled to PMTs. The pulse from one of the PMTs is used as a trigger, while the pulse from the other one is used for the energy and pulse measurement. The spectrum data is obtained using a CAEN DT5720 digitizer. The pulse-data is collected using a digital oscilloscope (Tektronix DPO4104) which is capable of storing data through USB port in a thumb-drive. \\

\begin{figure}[htbp!]
    \begin{subfigure}{0.5\textwidth}	
		\centering
		\includegraphics[height=55mm]{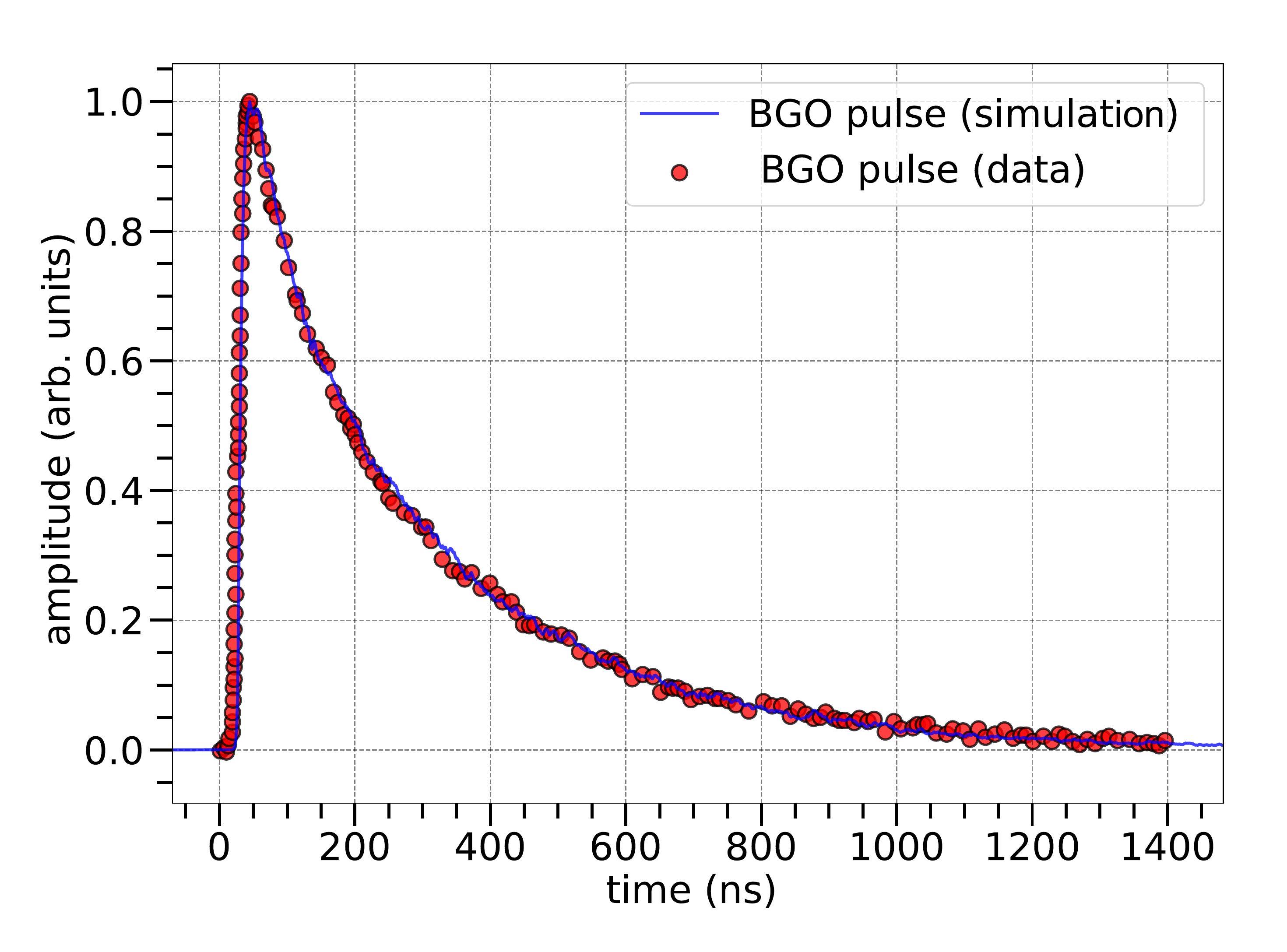}		
		\caption{}
		\label{BGOpulse} 
	\end{subfigure}%
	\begin{subfigure}{0.5\textwidth}
		\centering
		\includegraphics[height=50mm]{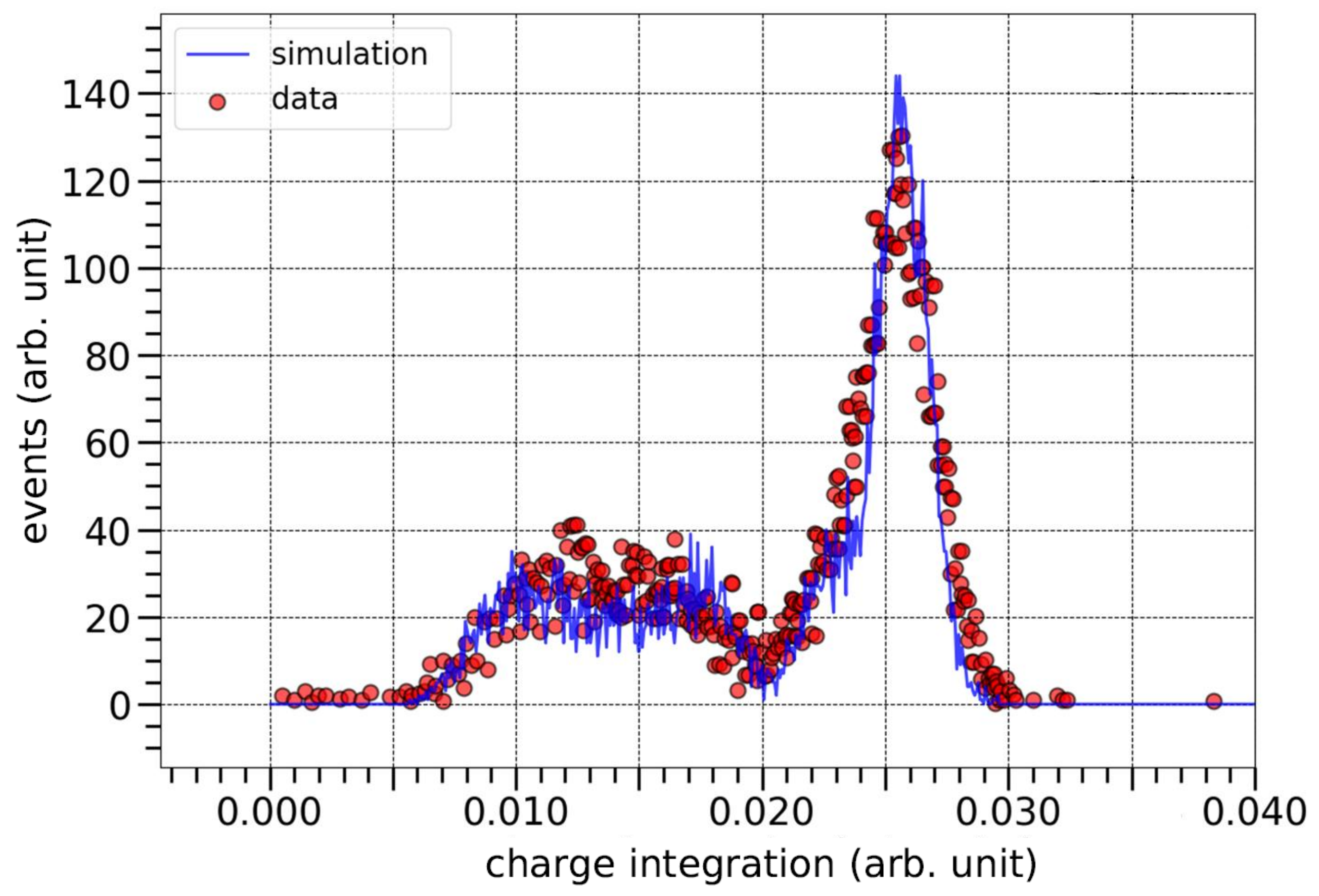}
		\caption{}
		\label{BGOspectra} 
	\end{subfigure}
	\caption{Simulation for BGO. a) Pulse shape comparison by taking average of 100 pulses. b) $\gamma$-spectra with $662$ keV $^{137}$Cs source measured using BGO scintillator. Both the plots show good agreement with experimental data.}
	\label{simBGO}
\end{figure}

\begin{figure}[htbp!]
    \begin{subfigure}{0.5\textwidth}	
		\centering
		\includegraphics[height=55mm]{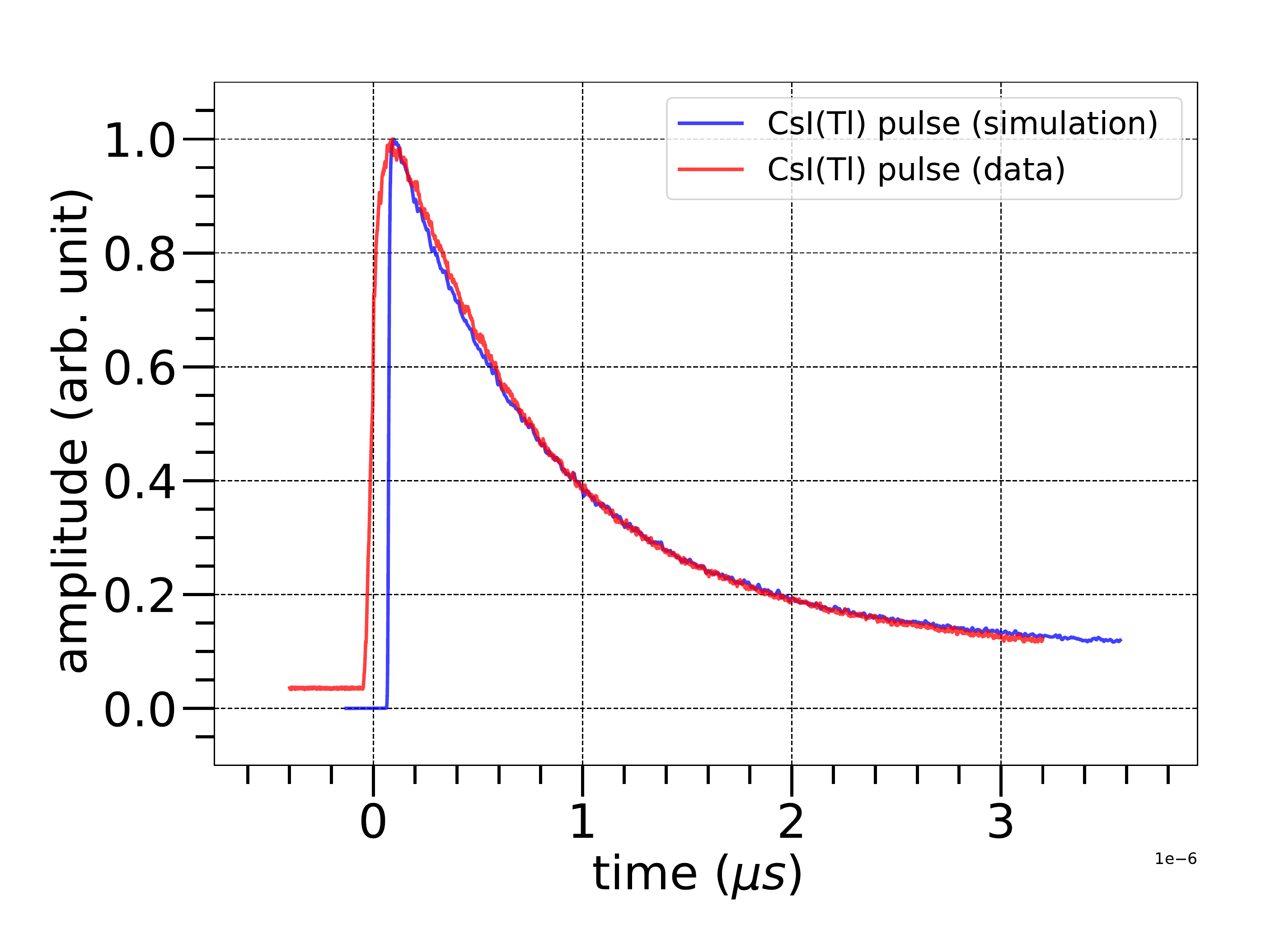}		
		\caption{}
		\label{CsIpulse} 
	\end{subfigure}%
	\begin{subfigure}{0.5\textwidth}
		\centering
		\includegraphics[height=50mm]{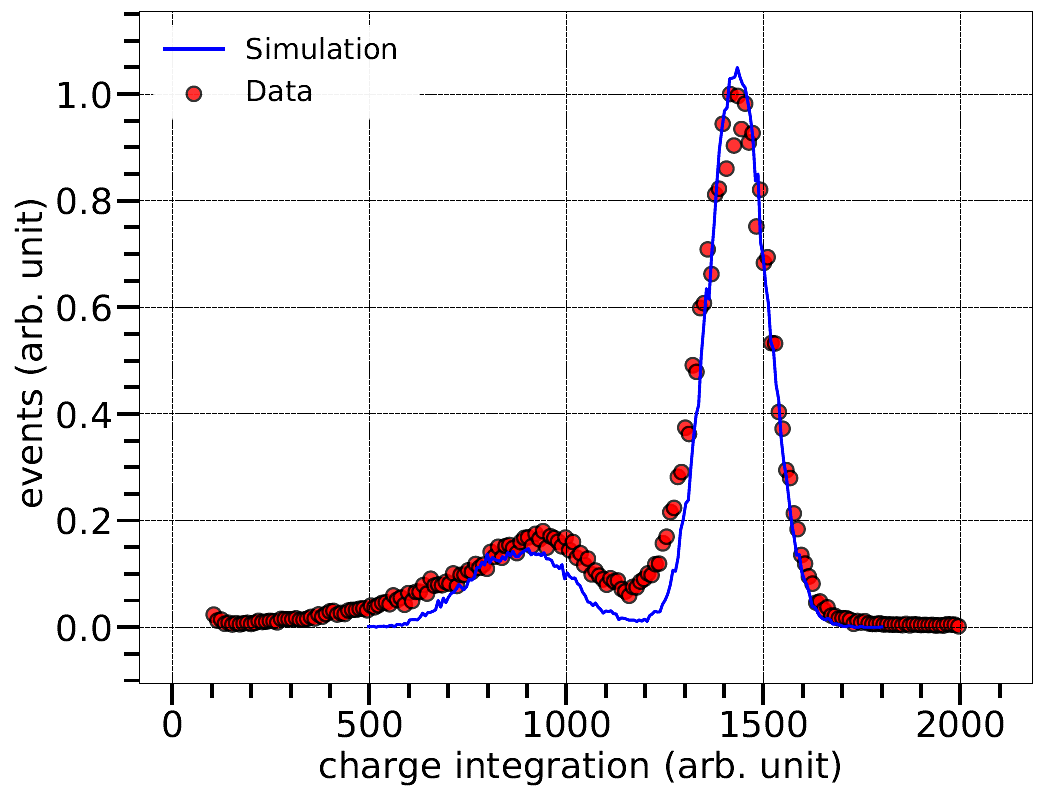}
		\caption{}
		\label{CsIspectra} 
	\end{subfigure}
	\caption{Simulation for CsI(Tl). a) Pulse shape comparison by taking average of 100 pulses. b) $\gamma$-spectra with $662$ keV $^{137}$Cs source measured using CsI(Tl) scintillator. The experimentally obtained spectra is broader than the simulated one. This can be attributed to the presence of electronic noise in the experimental data.}
	\label{simCsI}
\end{figure}

\noindent 
The validation of the simulation was done in two ways. First, by taking average of 100 pulses near the photo-peak, when the scintillators are irradiated with the $^{137}$Cs source and comparing the same with simulation. In order to validate the variation of the pulse due to difference in energy deposited within the crystal, comparison with experimentally obtained energy spectra is done. The simulated energy spectra was obtained by time-integrating the simulated pulses. The results are shown in figure \ref{simBGO} and \ref{simCsI} for BGO and CsI(Tl) respectively.
\noindent
The spectra matches reasonably well near the photo-peak and Compton valley in both the cases. The experimental data for BGO has a backscatter peak which is less prominent in the simulation. In case of CsI, the Compton valley is more broadened in the experimental data due to the presence of electronic noise. \\

\noindent
As mentioned in section \ref{introduction}, the pulses for different particles differ by their decay time for the same energy deposit \cite{Lee2014}. In order to study the PSD capabilities of the different methods explored in this paper, two categories of pulses were simulated that differ by their decay times. Both the categories of pulses are generated by impinging $\gamma$-ray photons on the scintillators. The first category of pulses, labelled as signal, have the decay times obtained from literature and/or experiment. The second category of pulses, labelled as background, have decay-time components (both fast and slow) 10\% larger than the first category. The scintillation parameters provided to GEANT4 for the two categrories are summarized in Table \ref{scintillationParameter}. The labelling as signal and background for the two categories have been used for the sake of clarity in the text when referring to them. \\

\begin{table}[ht!]
\begin{center}
    \begin{tabular}{|c| c| c|}
    \hline
    Parameter & Category I (signal) & Category II (background) \\
    \hline
    Rise Time ($ns$) & $296$ & $296$ \\
    Fast Decay Time ($\mu s$) & 1.30 & 1.43 \\
    Slow Decay Time ($\mu s$) & 4.50 & 4.95 \\
    Light yield (per MeV) & $54000$ & $54000$ \\ 
    \hline
    \end{tabular}
    \caption{Scintillation parameter values given to GEANT4}
    \label{scintillationParameter}
\end{center}
\end{table}

\noindent
Since, the motivation is to have better PSD capabilities at low energy deposits (10s of keV) in the scintillator, the simulated pulses for both signal and background are obtained by allowing low energy $\gamma$-rays to fall on the scintillators. The low energy analysis is done for 4 keV and 20 keV incident $\gamma$-rays. 50K samples were generated for each category (signal and background) for a total of 100K samples out of which 70K samples were used for training and 30K for testing. The samples for training and testing were selected randomly.

\section{Analysis}
\label{analysis}

\noindent
The performance of the ML algorithms have been compared with conventional methods. All the methods were applied to the same signal and background datasets and their discrimination capability is analyzed. The figure-of-merit is obtained by making Receiver Operating Characteristic (ROC) plots and then calculating the area-under-curve (AUC). The ROC curve is made by calculating the signal and background efficiencies, defined as fraction of signal and background events that pass a particular selection criteria respectively, and then plotting these quantities in the background efficiency vs. signal efficiency plane for all selection criteria.\\

\begin{figure}[ht!]
    \begin{subfigure}{0.33333\textwidth}	
		\centering
		\includegraphics[height=50mm]{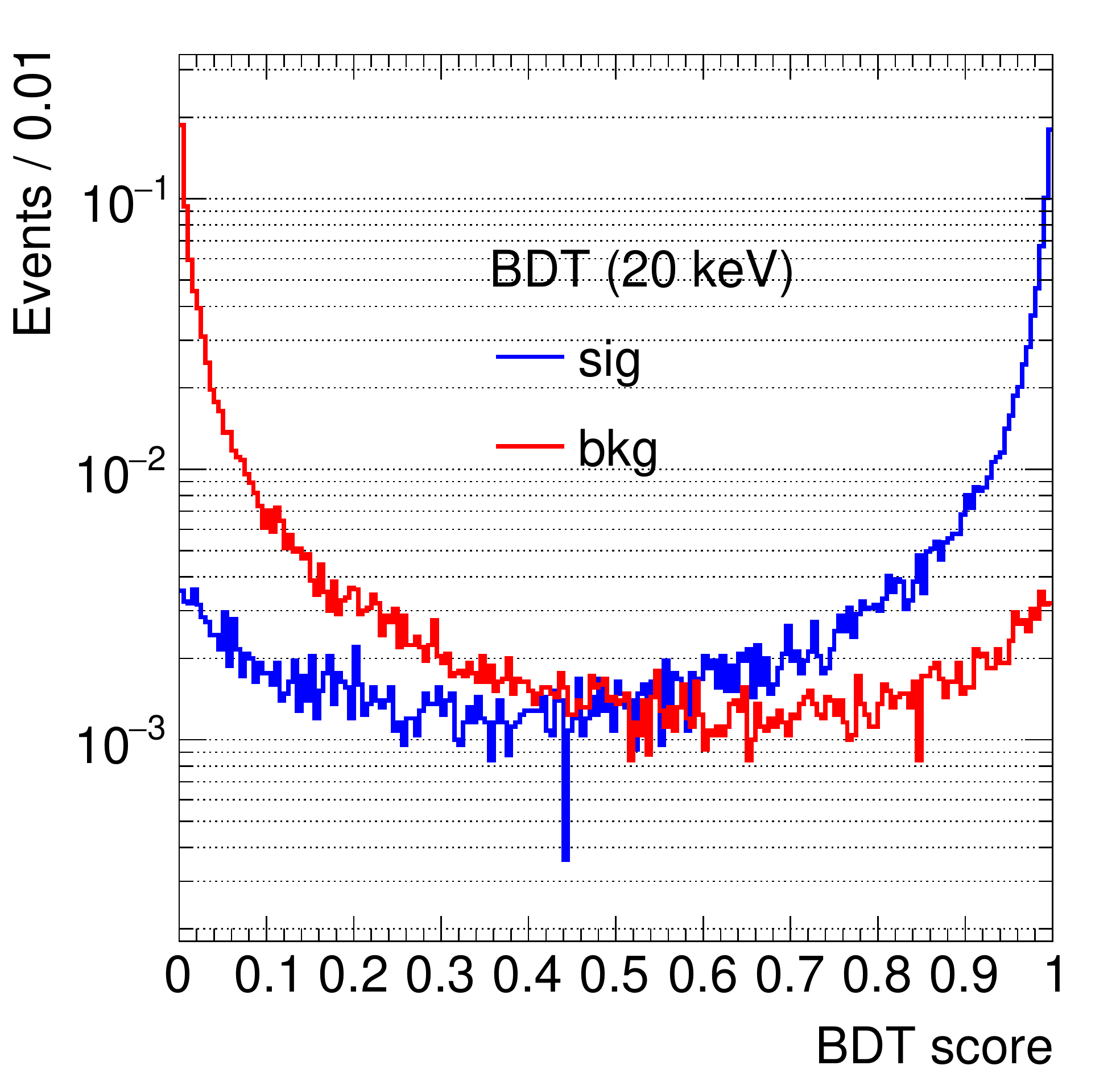}	
		\caption{}
		\label{scoreBDT_20keV} 
	\end{subfigure}%
	\begin{subfigure}{0.33333\textwidth}	
		\centering
		\includegraphics[height=50mm]{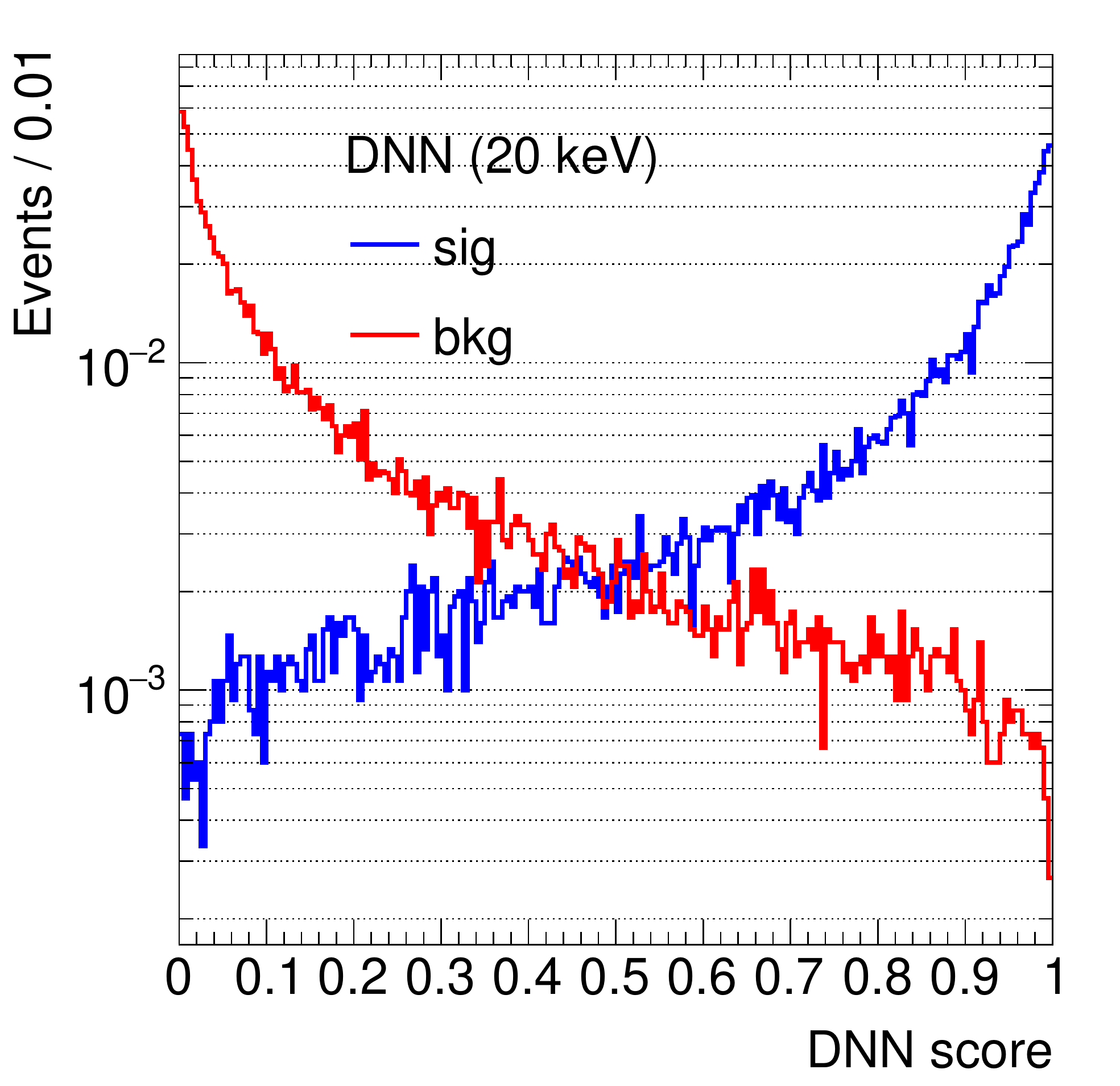}	
		\caption{}
		\label{scoreDNN_20keV} 
	\end{subfigure}%
	\begin{subfigure}{0.33333\textwidth}	
		\centering
		\includegraphics[height=50mm]{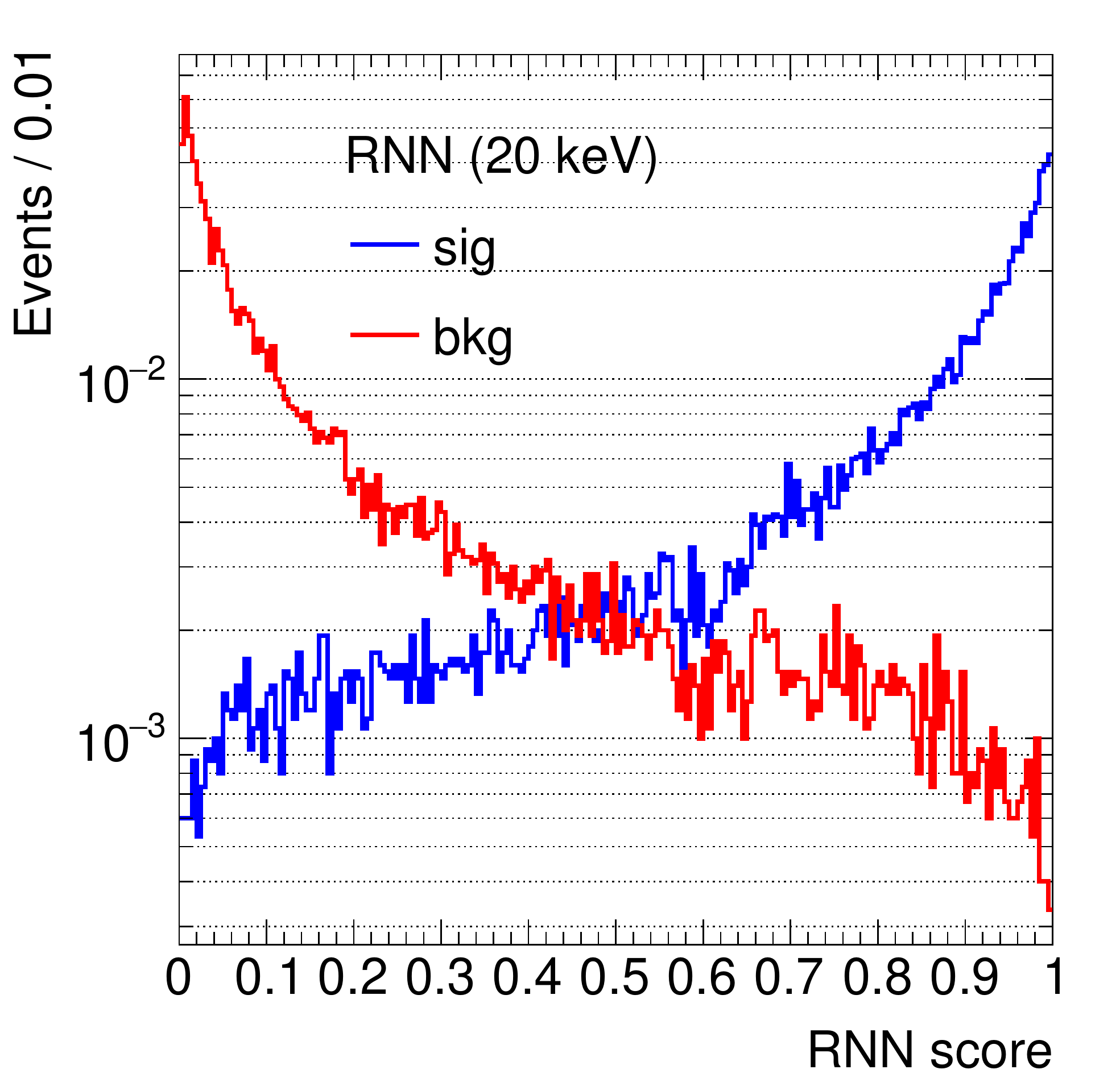}	
		\caption{}
		\label{scoreRNN_20keV} 
	\end{subfigure}	
    \caption{Score distributions of ML methods for 20 keV recoil energy. a) BDT. b) DNN. c) RNN.}
	\label{MLscore_20keV}
\end{figure}

\begin{figure}[ht!]
    \begin{subfigure}{0.33333\textwidth}
		\centering
		\includegraphics[height=50mm]{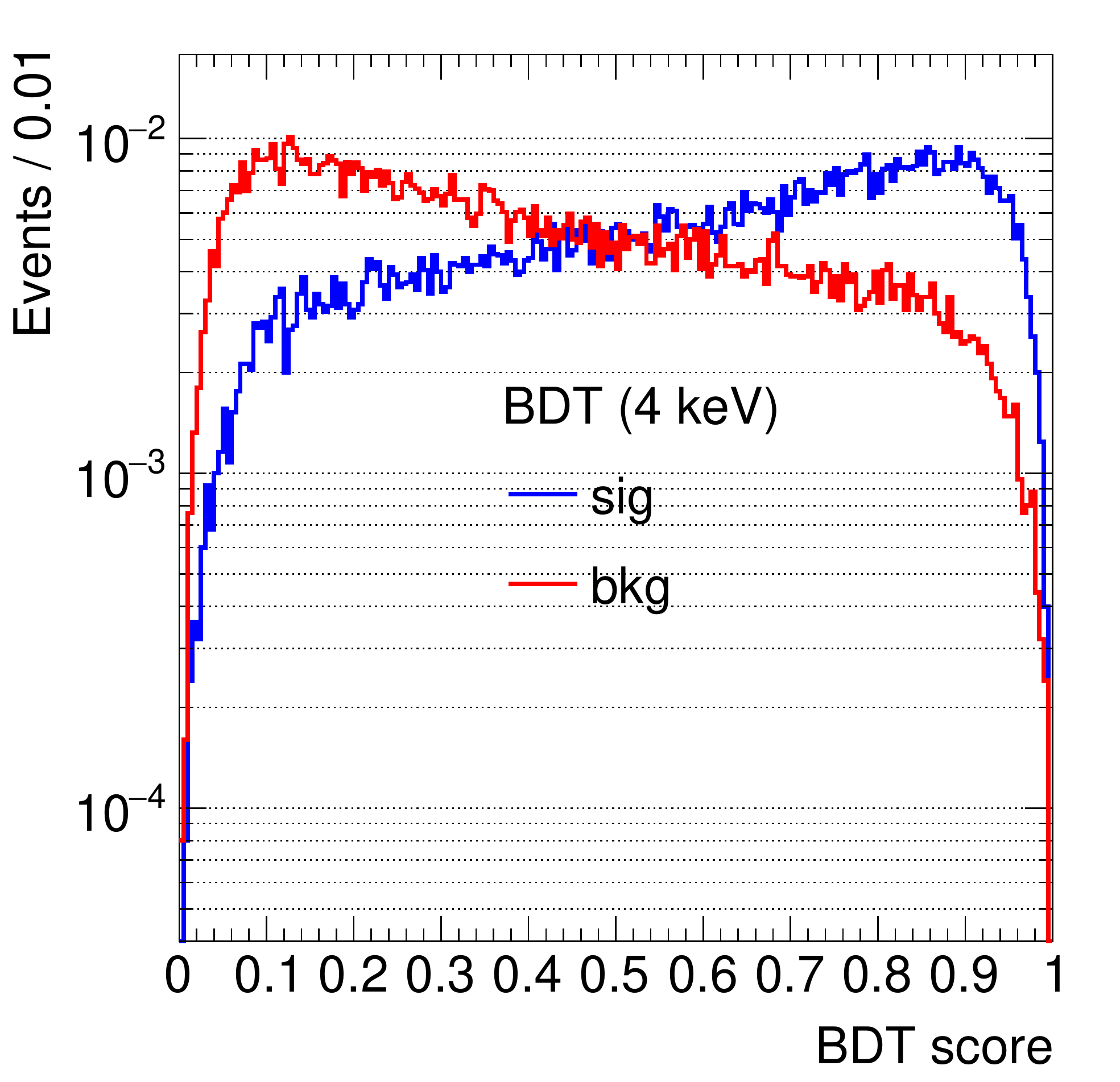}
		\caption{}
		\label{scoreBDT_4keV} 
	\end{subfigure}%
	\begin{subfigure}{0.33333\textwidth}
		\centering
		\includegraphics[height=50mm]{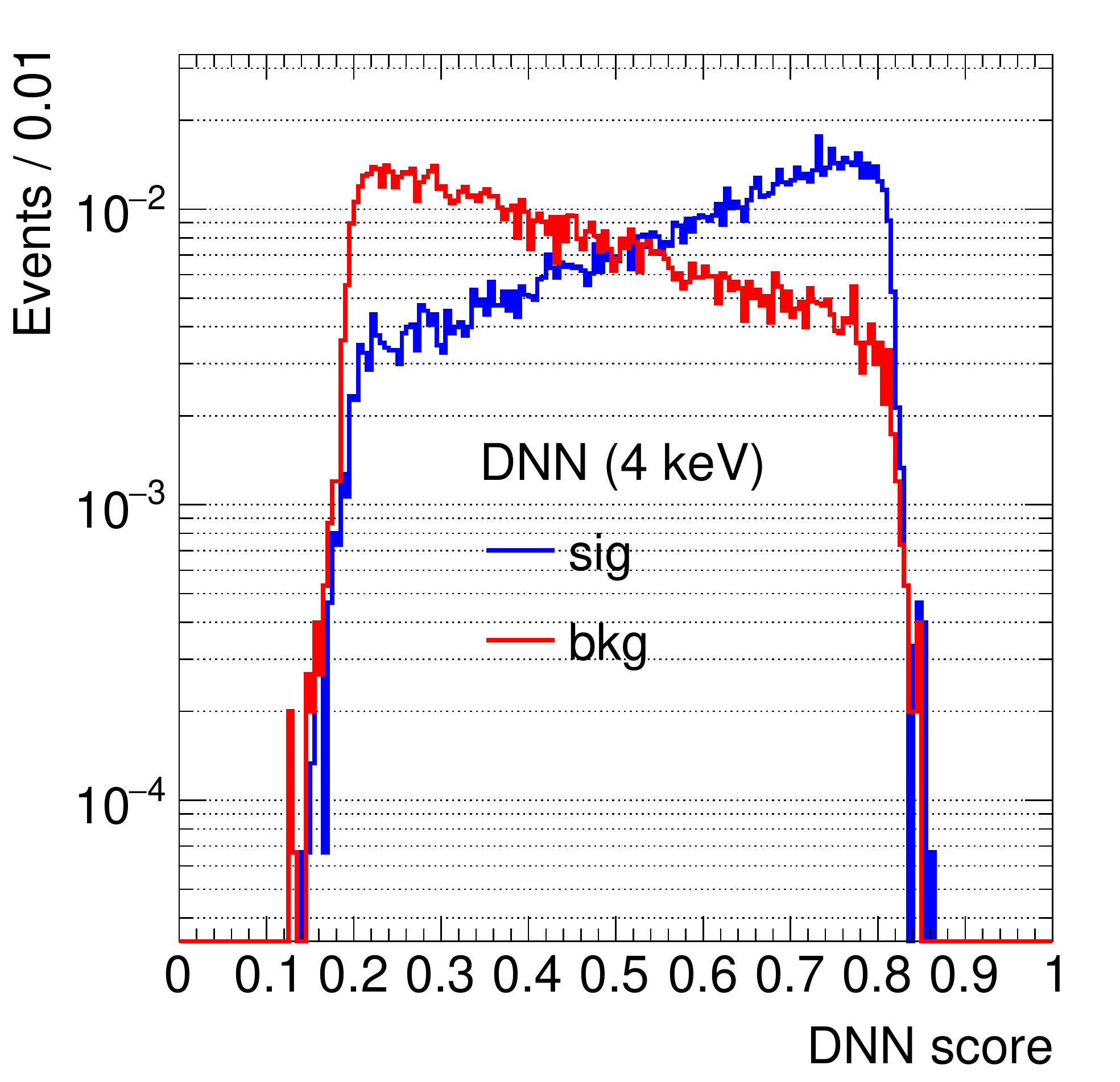}
		\caption{}
		\label{scoreDNN_4keV} 
	\end{subfigure}%
	\begin{subfigure}{0.33333\textwidth}
		\centering
		\includegraphics[height=50mm]{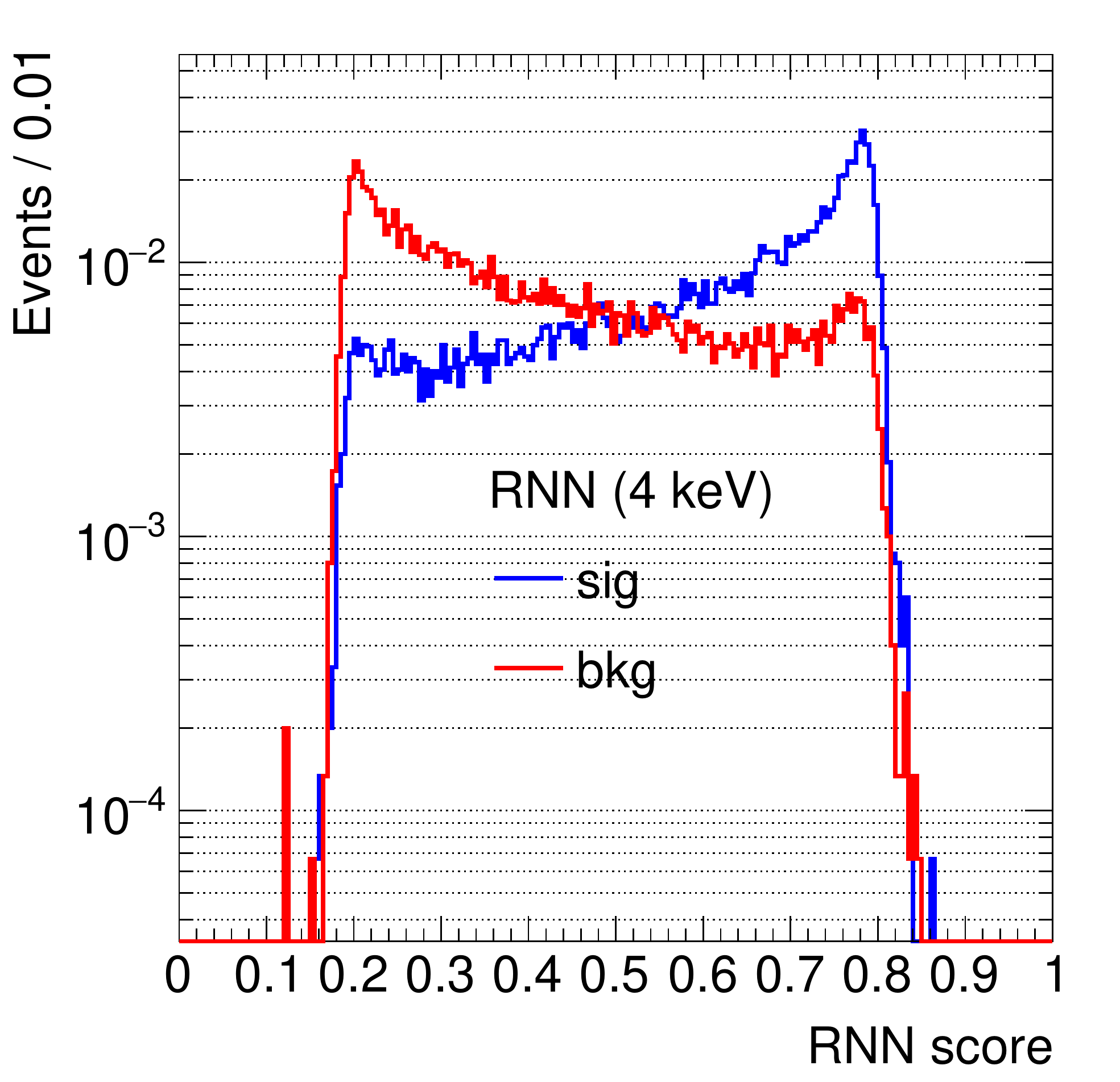}
		\caption{}
		\label{scoreRNN_4keV} 
	\end{subfigure}%
    \caption{Score distributions of ML methods for 4 keV recoil energy. a) BDT. b) DNN. c) RNN.}
	\label{MLscore_4keV}
\end{figure}

\begin{figure}[ht!]
    \begin{subfigure}{0.5\textwidth}	
		\centering
		\includegraphics[height=70mm]{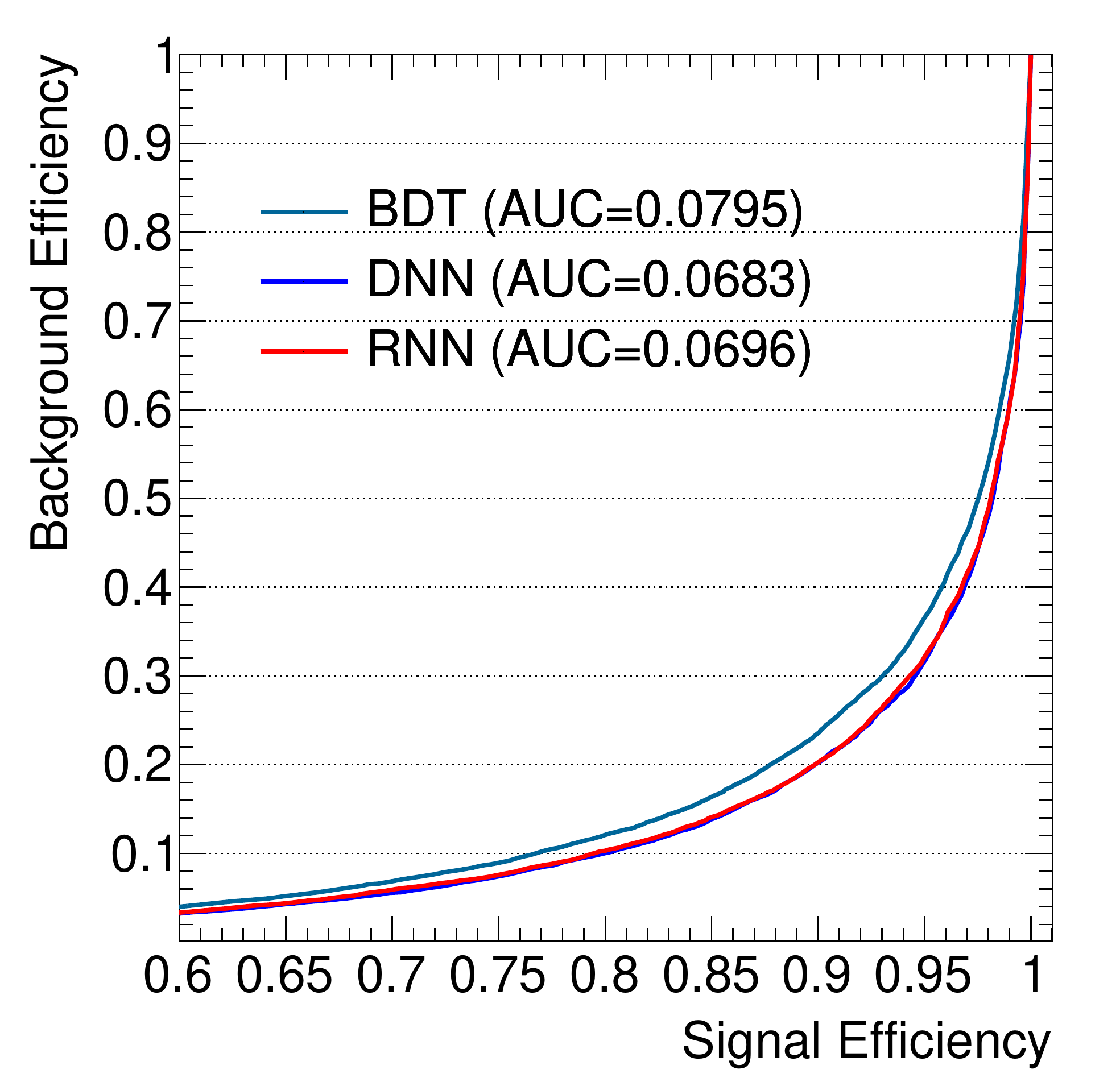}	
		\caption{}
		\label{ROC_BDT_NN_20keV} 
	\end{subfigure}%
	\begin{subfigure}{0.5\textwidth}
		\centering
		\includegraphics[height=70mm]{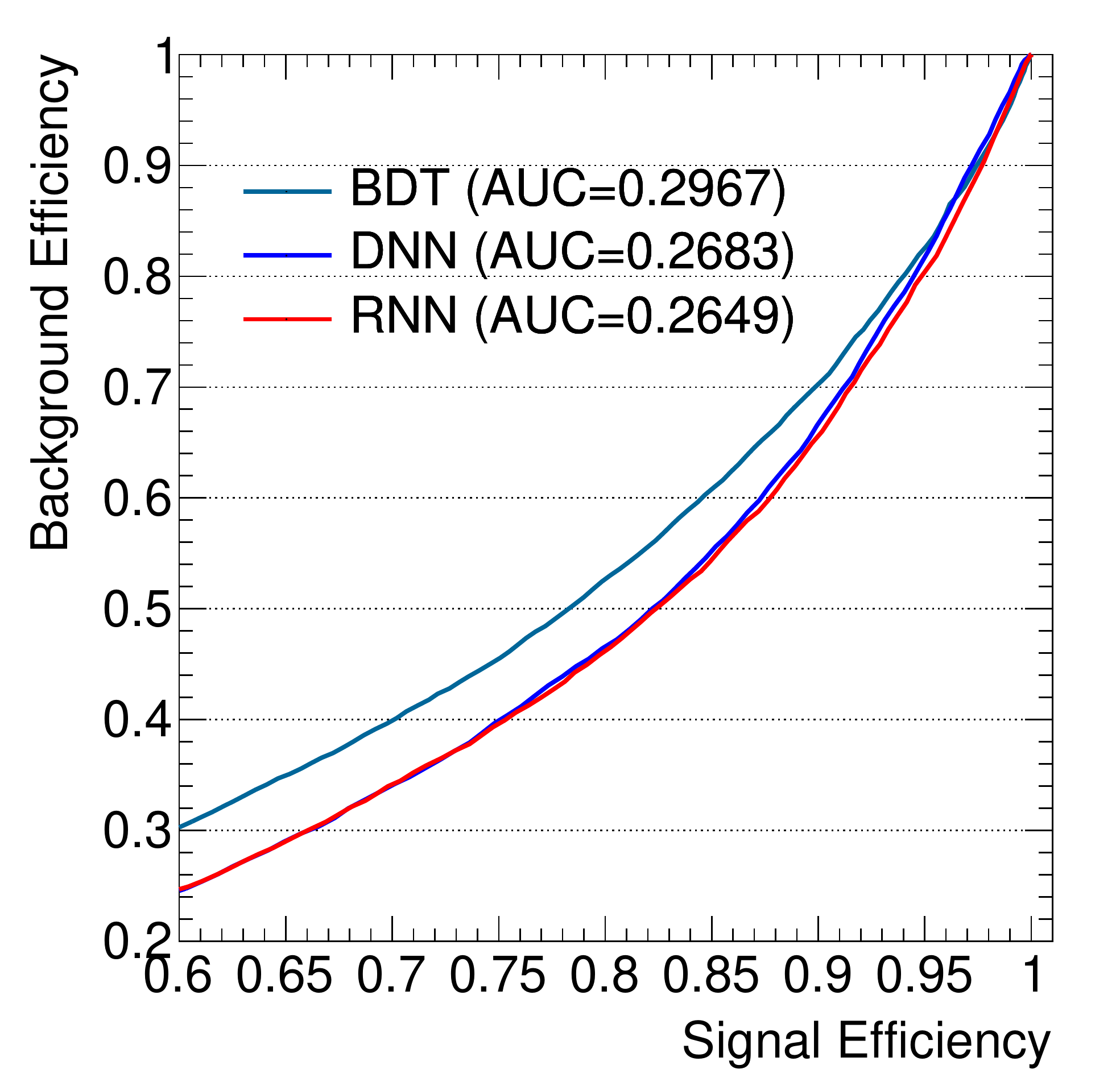}
		\caption{}
		\label{ROC_BDT_NN_4keV} 
	\end{subfigure}
    \caption{ROC plots comparing the ML methods for a) 20 keV and b) 4 keV. The area-under-curve (AUC) values are indicated in the legends.}
	\label{ROC_BDT_NN}
\end{figure}

\noindent
Figures \ref{MLscore_20keV} and \ref{MLscore_4keV} show the distribution of the scores of the ML methods at recoil energies of 20 keV and 4 keV respectively. The ROC plots comparing the performance of the different ML methods for 20 keV and 4 keV are shown in figures \ref{ROC_BDT_NN_20keV} and \ref{ROC_BDT_NN_4keV}; the AUC values are mentioned alongside the legends. It can be concluded from the AUC values, that the network-based methods perform better than BDT. So, only the network based methods have been compared with the conventional methods of Charge Integration (CI) and Mean-time (lnMT) method. \\ 

\begin{figure}[htbp!]
   	\centering
	\includegraphics[height=70mm]{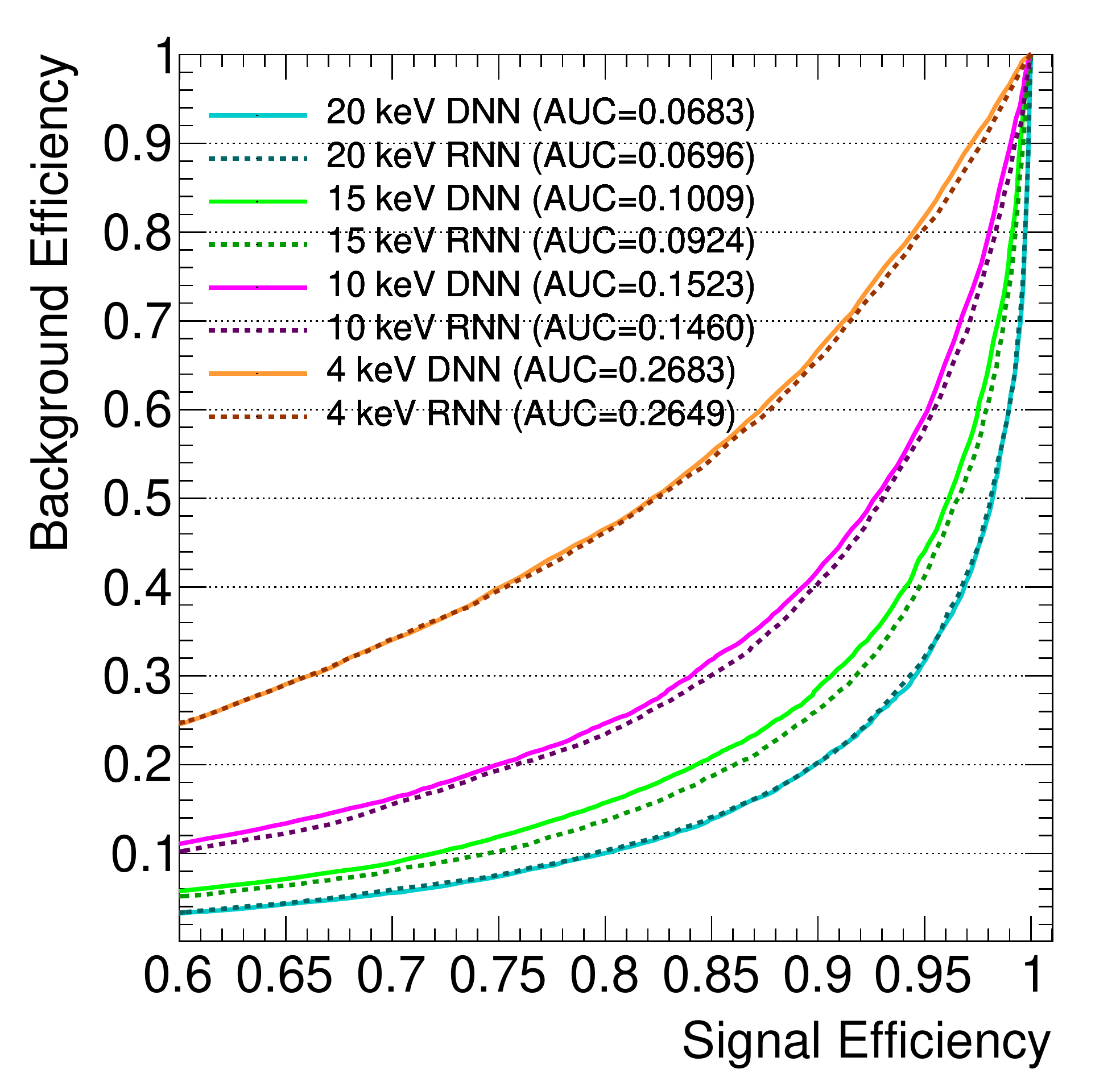}
	\caption{ROC curves corresponding to 20 keV, 15 keV, 10 keV and 4 keV recoil energies for DNN and RNN as labelled. This clearly shows that the performance degrades as we go to lower energies.}
	\label{ROC_DNN_RNN}
\end{figure}

\noindent
The ROC plots in figure \ref{ROC_BDT_NN} show that the performance degrades as we go to lower energy. This is also indicated by the broadening of the score distribution in case of 4 keV (figure \ref{MLscore_4keV}) compared to 20 keV (figure \ref{MLscore_20keV}). In order to have a clearer picture of this, we plotted the ROC for four energies (20 keV, 15 keV, 10 keV and 4 keV), each for DNN and RNN in figure \ref{ROC_DNN_RNN}. In all the cases, except for 20 keV, it can be seen that the RNN does marginally better than the DNN. \textcolor{blue}{This is also indicated by the training loss curves, which are given in Appendix \ref{App_lossCurves}. Thus,} the change in case of 20 keV can be attributed to statistical fluctuation.\\

\begin{figure}[ht!]
    \begin{subfigure}{0.5\textwidth}	
		\centering
		\includegraphics[height=60mm]{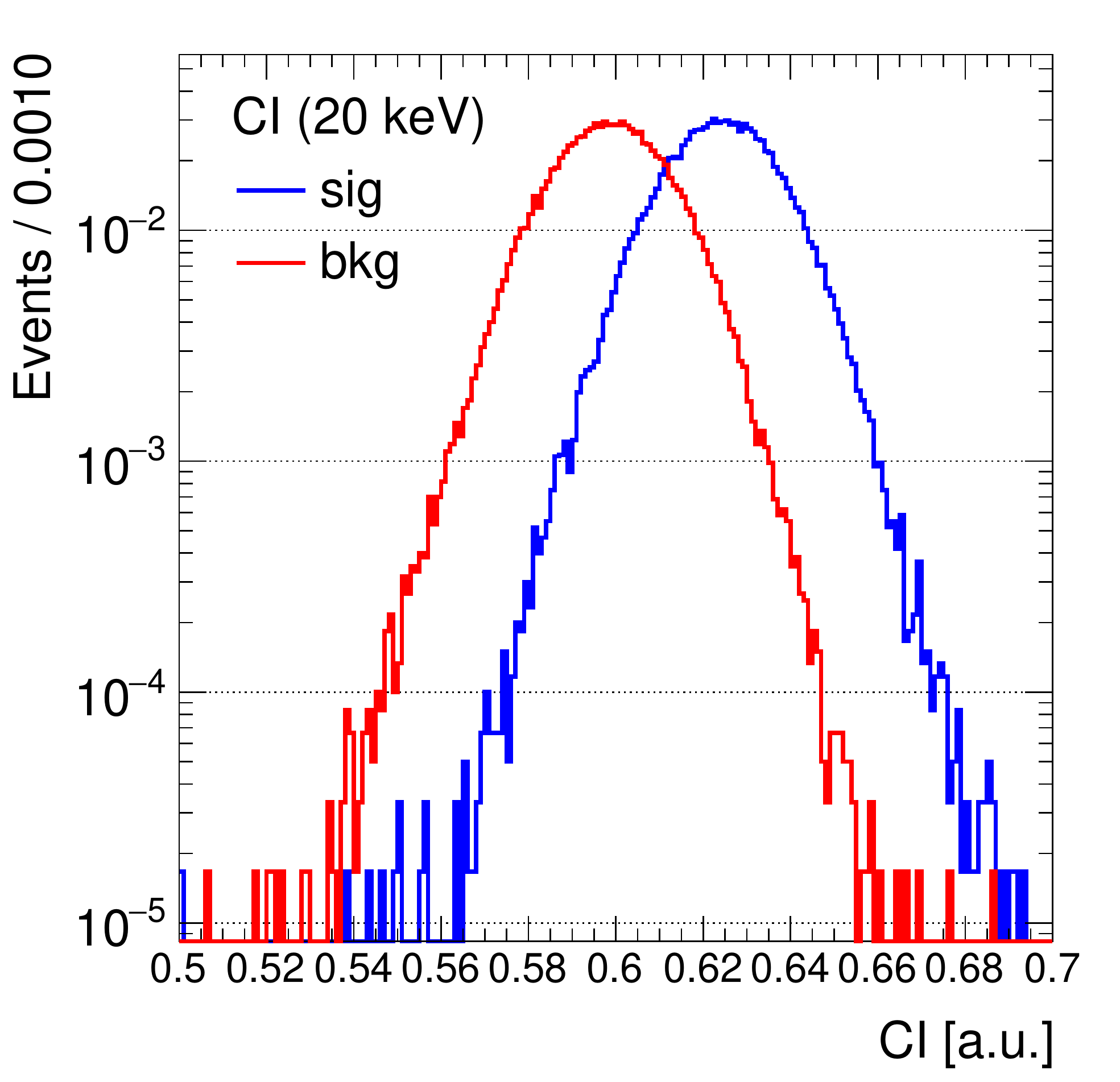}	
		\caption{}
		\label{SGbyLG_20keV_2} 
	\end{subfigure}%
	\begin{subfigure}{0.5\textwidth}
		\centering
		\includegraphics[height=60mm]{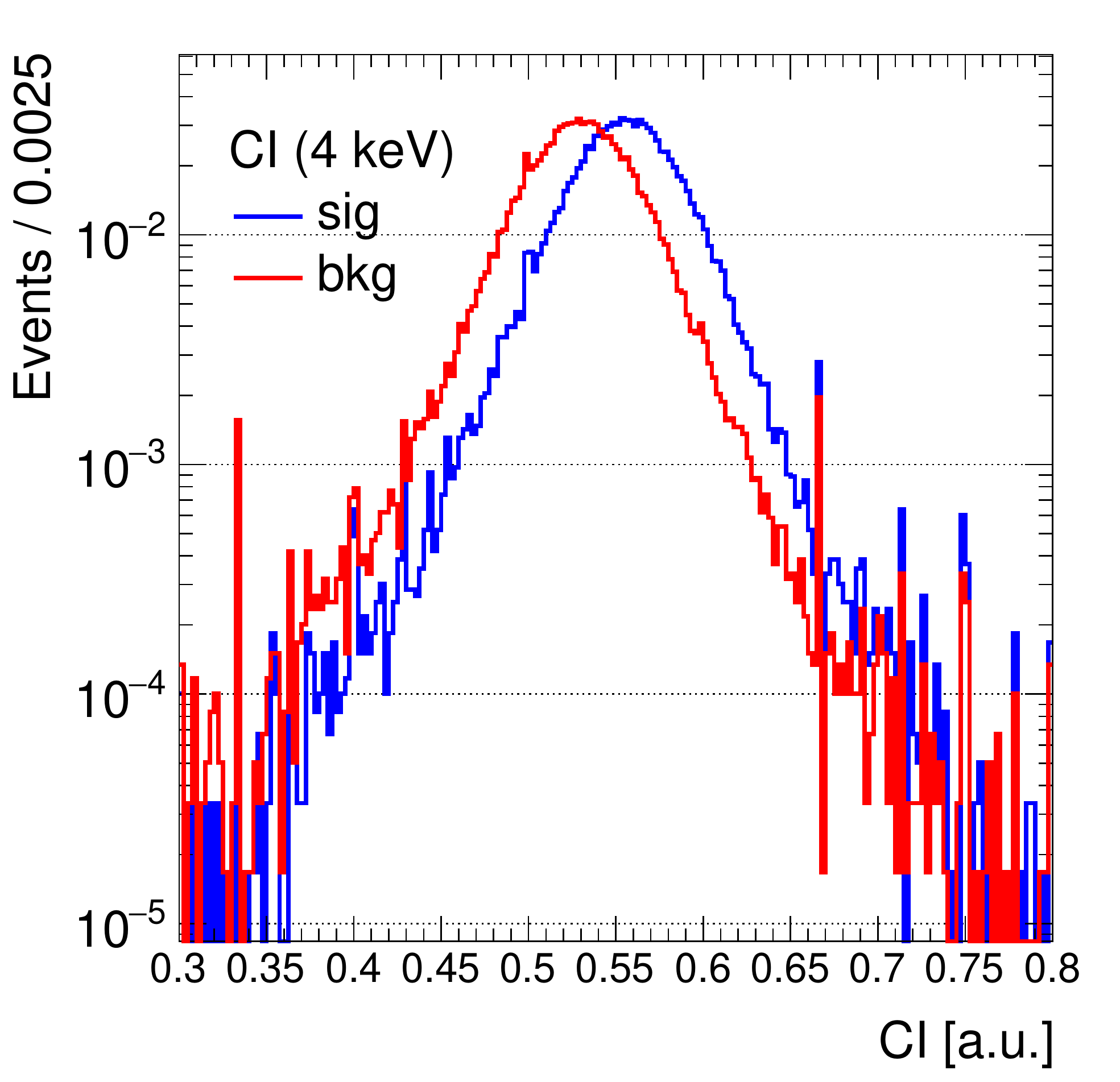}
		\caption{}
		\label{SGbyLG_4keV_2} 
	\end{subfigure}
	\caption{Distribution of integrated charge ratio for a) 20 keV and b) 4 keV}
	\label{SGbyLG}
\end{figure}

\begin{figure}[ht!]
    \begin{subfigure}{0.5\textwidth}	
		\centering
		\includegraphics[height=70mm]{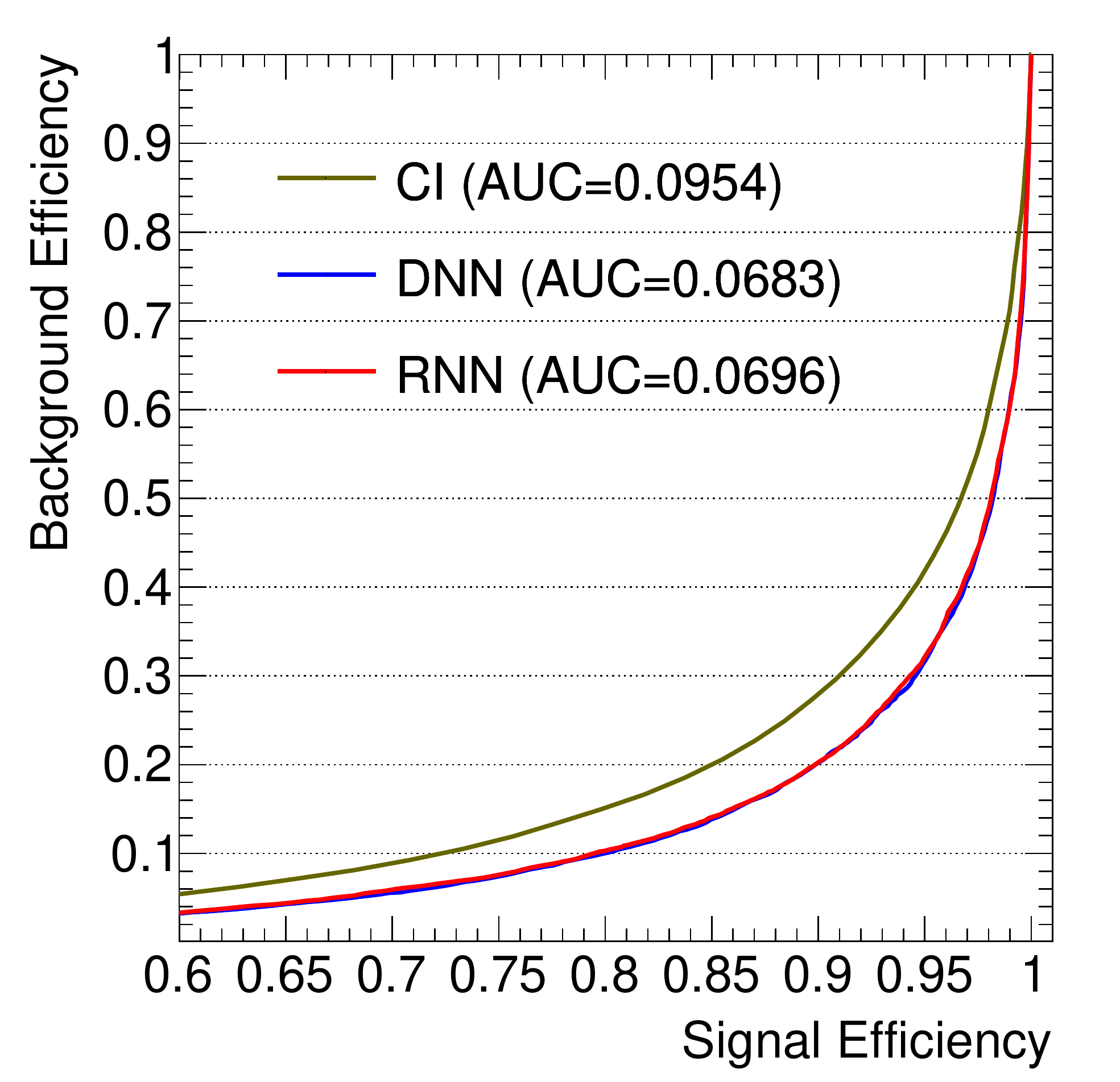}	
		\caption{}
		\label{ROC_SGbyLG_20keV} 
	\end{subfigure}%
	\begin{subfigure}{0.5\textwidth}
		\centering
		\includegraphics[height=70mm]{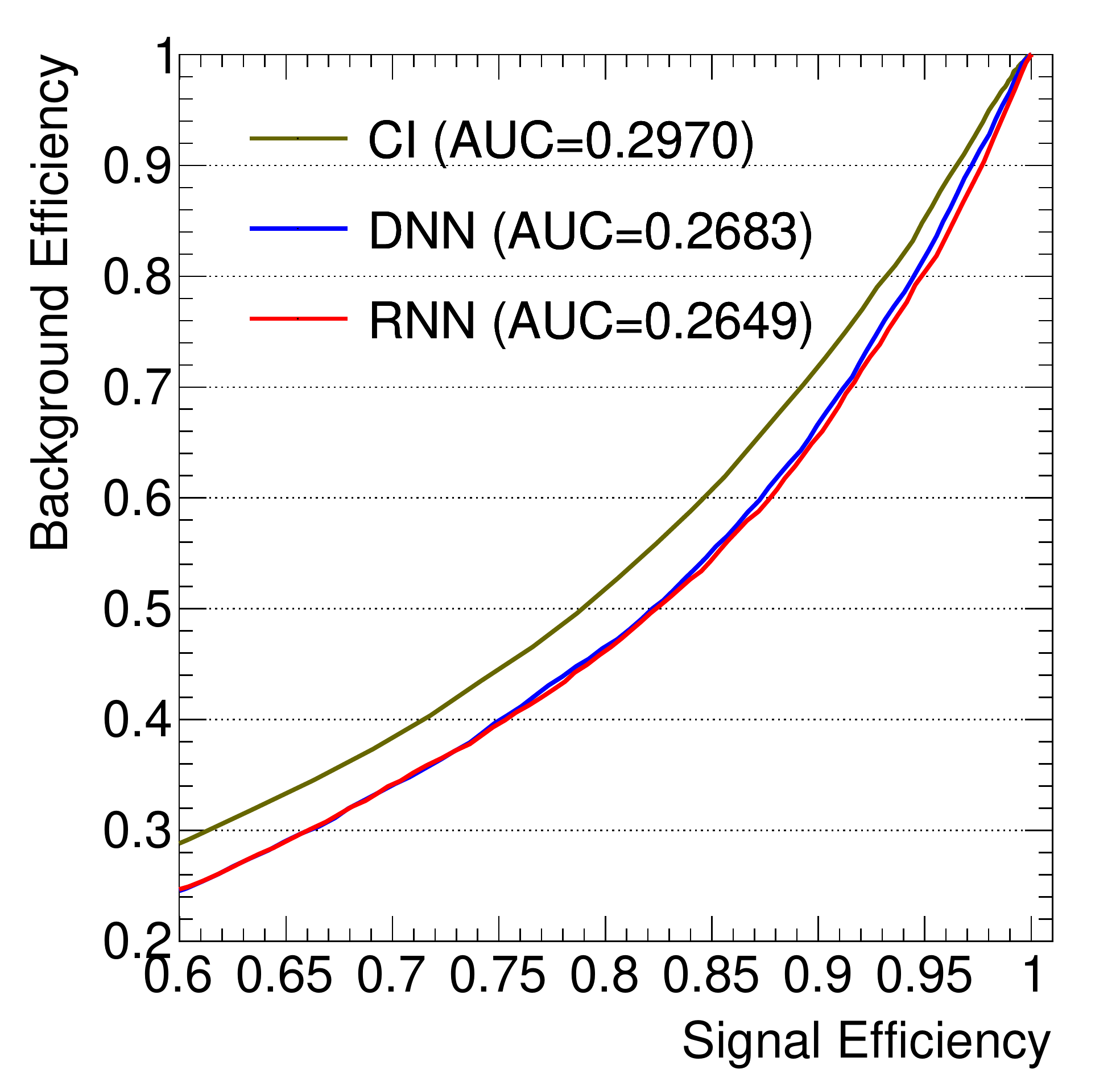}
		\caption{}
		\label{ROC_SGbyLG_4keV} 
	\end{subfigure}
    \caption{ROC plots comparing charge integration method with network for a) 20 keV and b) 4 keV. The network based methods have lower AUC values.}
	\label{ROC_SGbyLG_NN}
\end{figure}

\noindent
In the CI method, the ratio of charge within a short gate (time window) to the charge within a long gate is calculated. The distribution of the ratio is plotted in figure \ref{SGbyLG}. Different width of short gates were taken, but only the best one is shown. Figure \ref{ROC_SGbyLG_NN} shows the ROC plots for 20 keV and 4 keV along with the AUC values. The networks clearly provide better discrimination. \\  

\begin{figure}[ht!]
    \centering
    \begin{subfigure}{0.5\textwidth}
        \includegraphics[height=70mm]{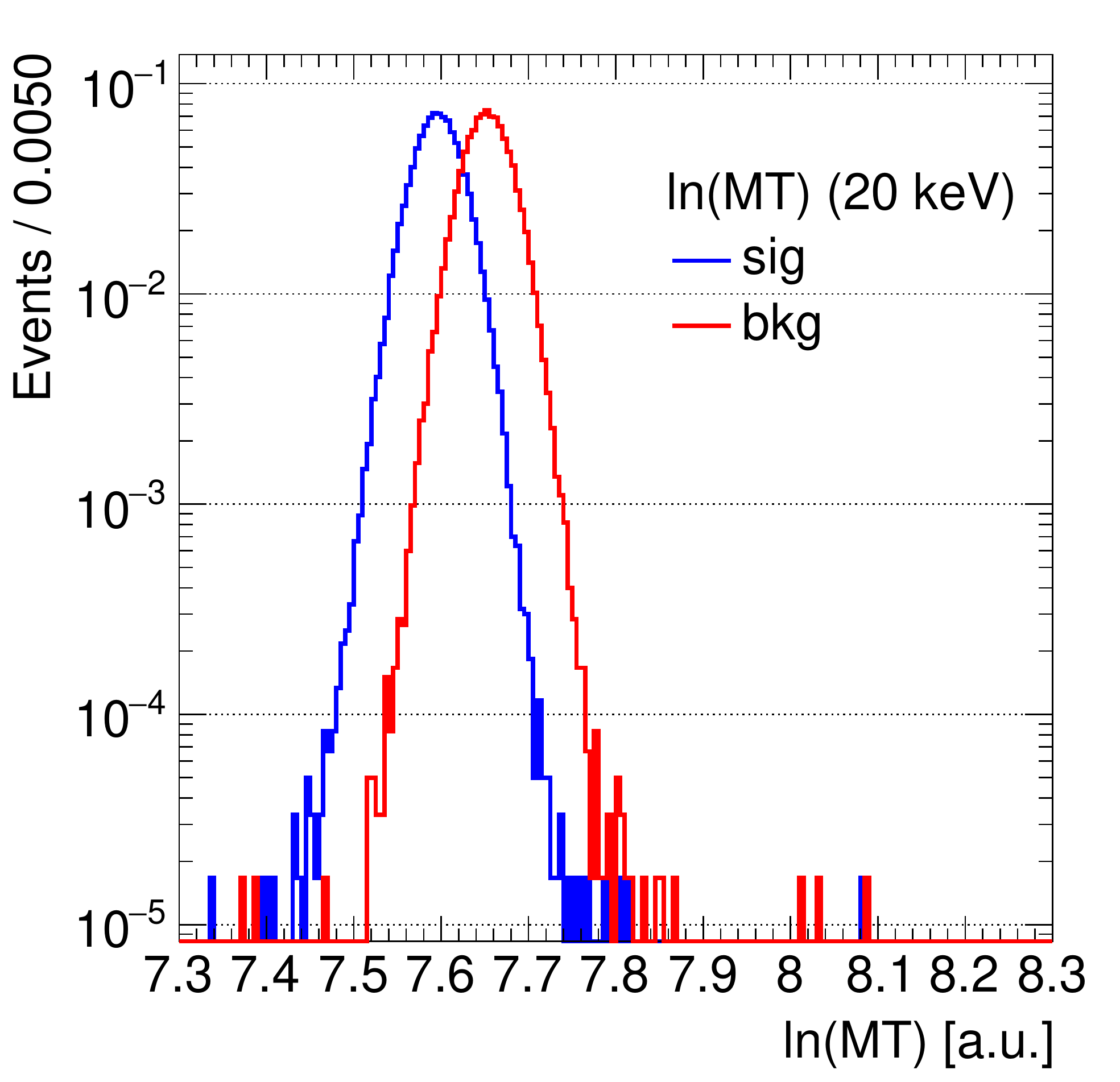}	
        \caption{}
        \label{lnMT_20keV}
    \end{subfigure}%
    \begin{subfigure}{0.5\textwidth}
        \includegraphics[height=70mm]{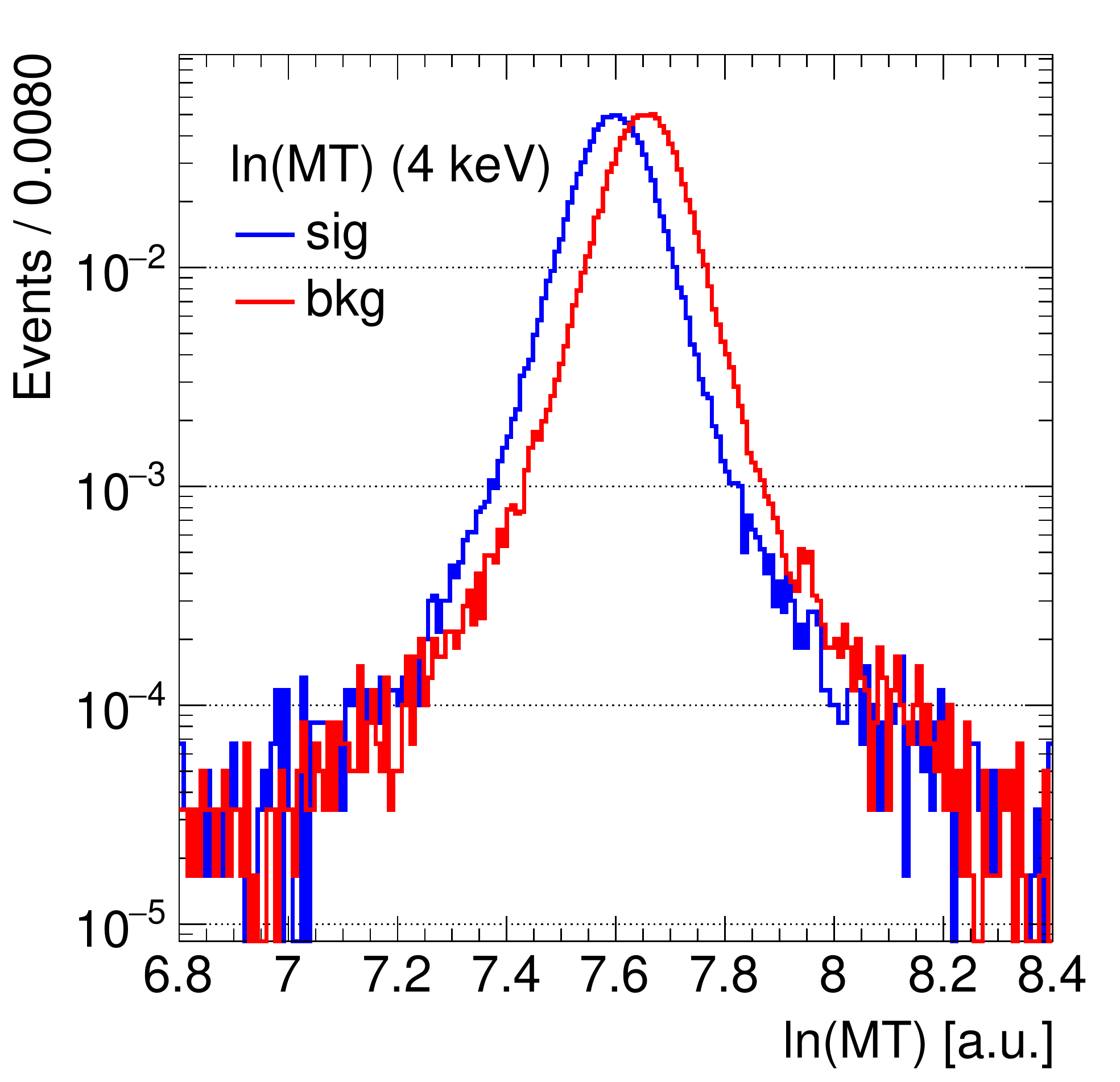}	
        \caption{}
        \label{lnMT_4keV}
    \end{subfigure}		
	\caption{Distribution of natural log of Mean Time. a) For 20 keV. b) For 4 keV.}
	\label{lnMT} 
\end{figure}

\begin{figure}[ht!]
    \begin{subfigure}{0.5\textwidth}	
		\centering
		\includegraphics[height=70mm]{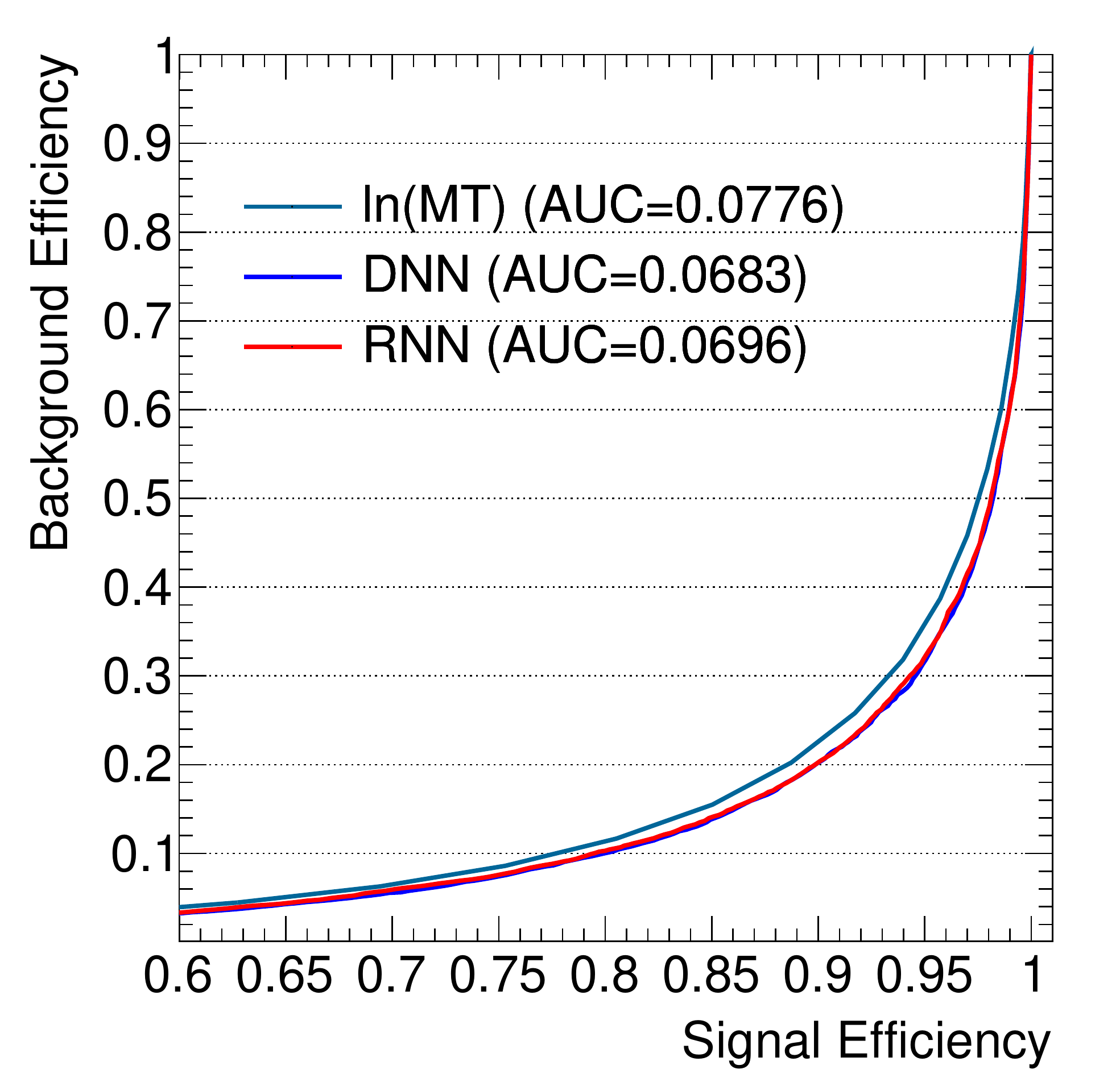}	
		\caption{}
		\label{ROC_lnMT_20keV} 
	\end{subfigure}%
	\begin{subfigure}{0.5\textwidth}
		\centering
		\includegraphics[height=70mm]{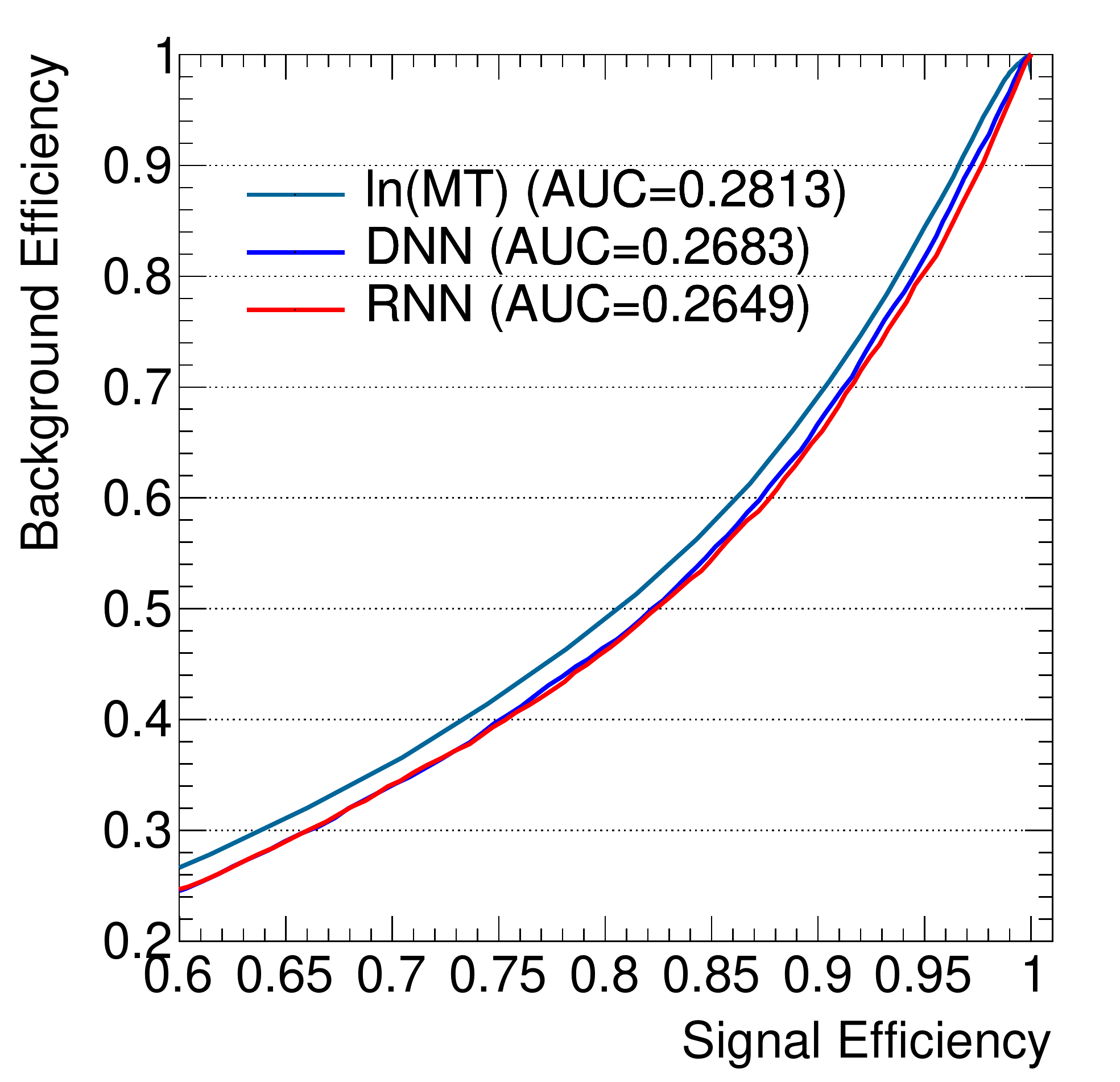}
		\caption{}
		\label{ROC_lnMT_4keV} 
	\end{subfigure}
    \caption{ROC plots comparing natural log of Mean-time with network for a) 20 keV and b) 4 keV. The network based methods show improvement over meant-time method.}
	\label{ROC_lnMT_NN}
\end{figure}

\begin{figure}[ht!]
    \begin{subfigure}{0.33333\textwidth}	
		\centering
		\includegraphics[height=50mm]{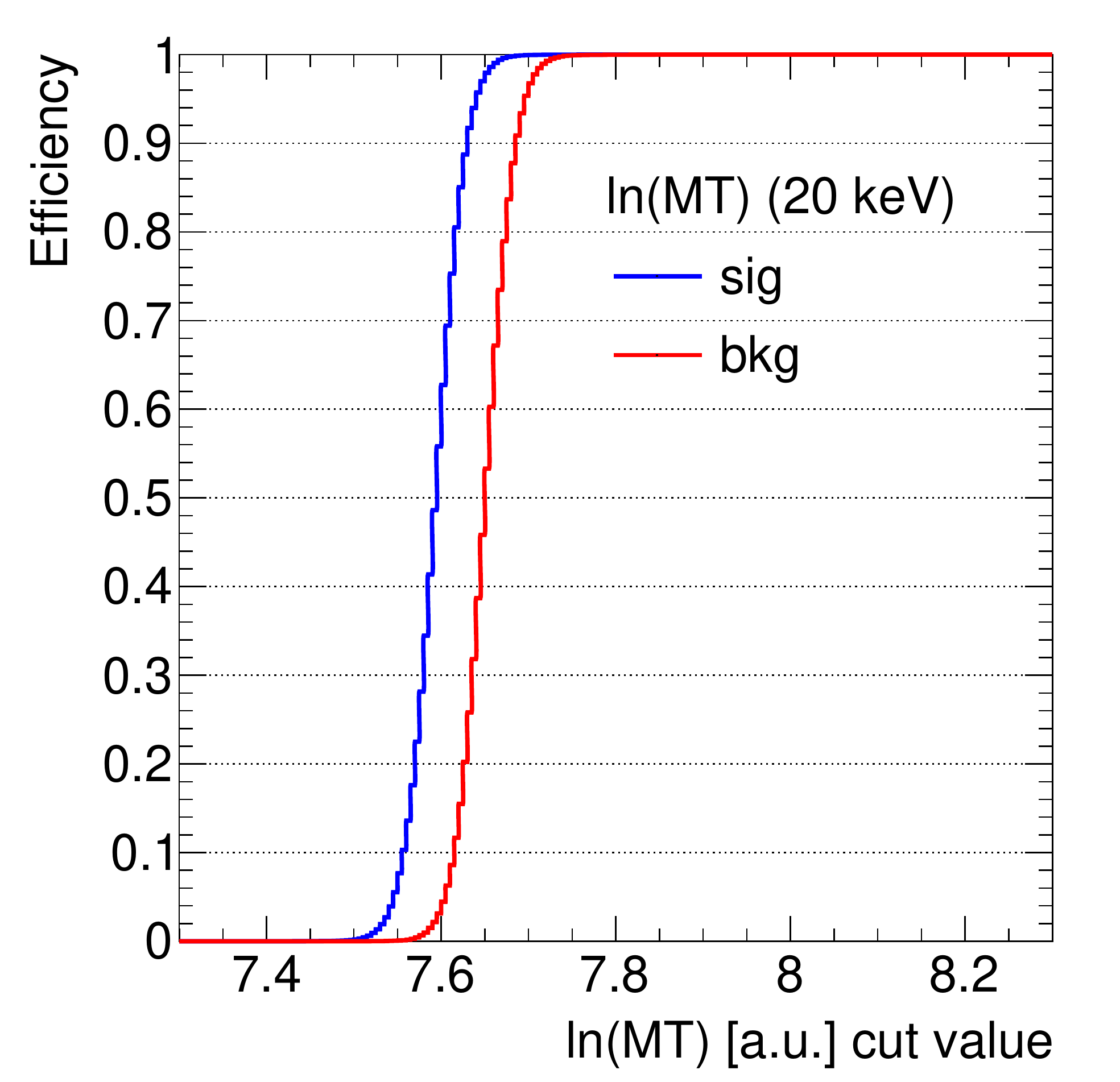}	
		\caption{}
		\label{sigVsbkgEff_lnMT_20keV} 
	\end{subfigure}%
	\begin{subfigure}{0.33333\textwidth}
		\centering
		\includegraphics[height=50mm]{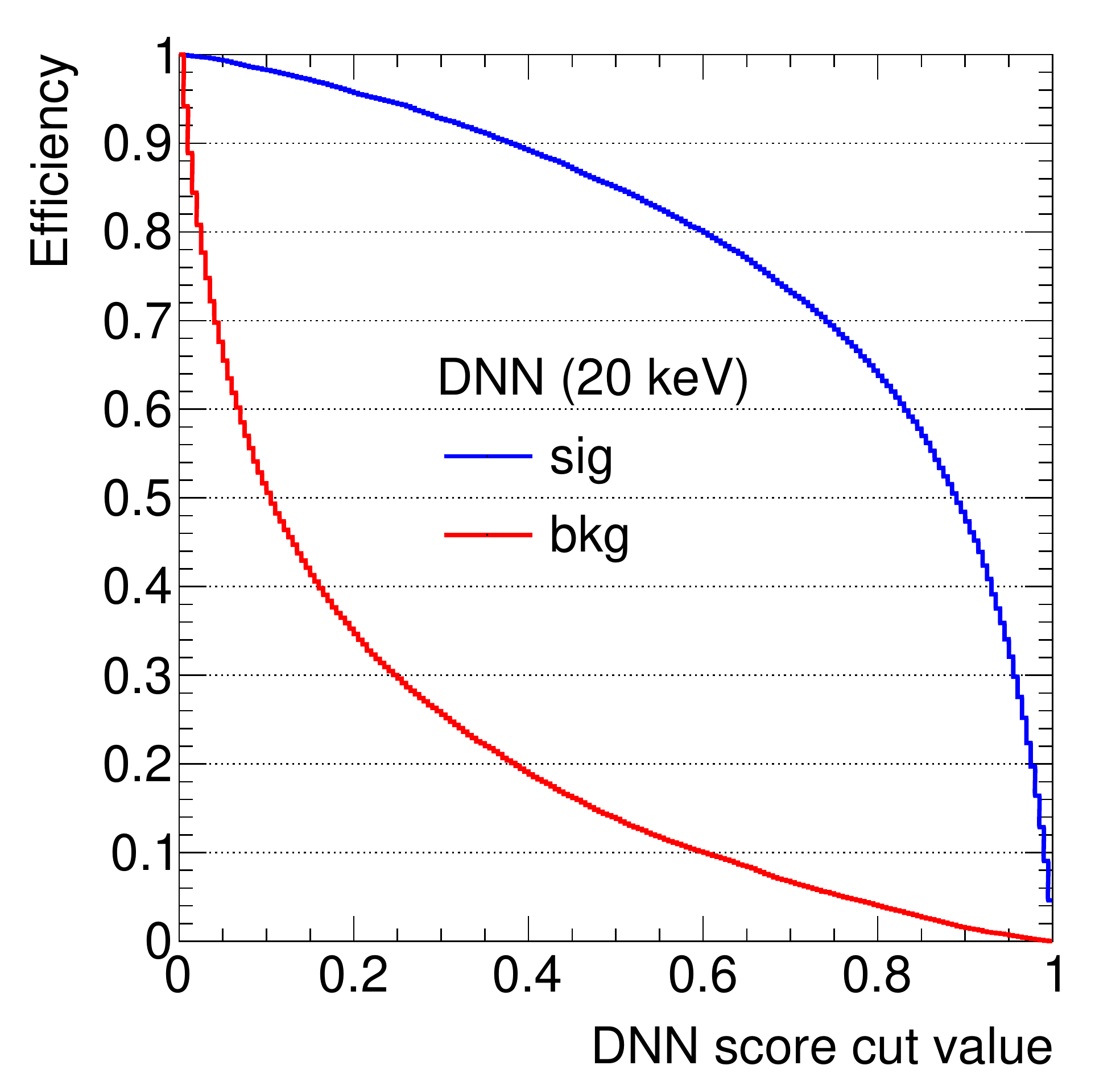}
		\caption{}
		\label{sigVsbkgEff_scoreDNN_20keV} 
	\end{subfigure}%
	\begin{subfigure}{0.33333\textwidth}
		\centering
		\includegraphics[height=50mm]{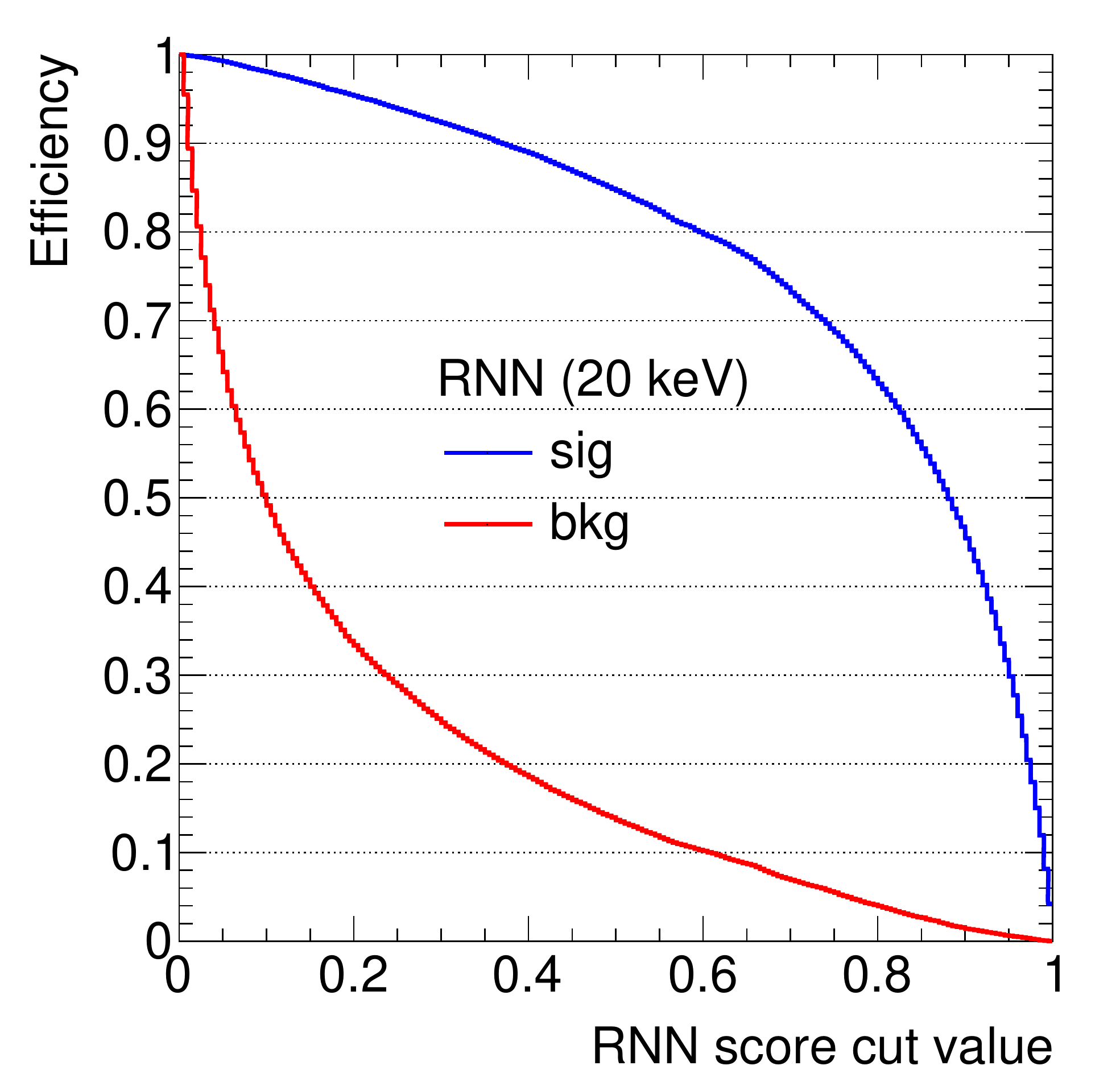}
		\caption{}
		\label{sigVsbkgEff_scoreRNN_20keV} 
	\end{subfigure}
    \caption{Signal and Background efficiency vs. Discriminating Threshold for 20 keV. a) Mean Time. b) DNN. c) RNN}
	\label{sigVsbkgEff_lnMT_NN_20keV}
\end{figure}

\begin{figure}[ht!]
    \begin{subfigure}{0.33333\textwidth}	
		\centering
		\includegraphics[height=50mm]{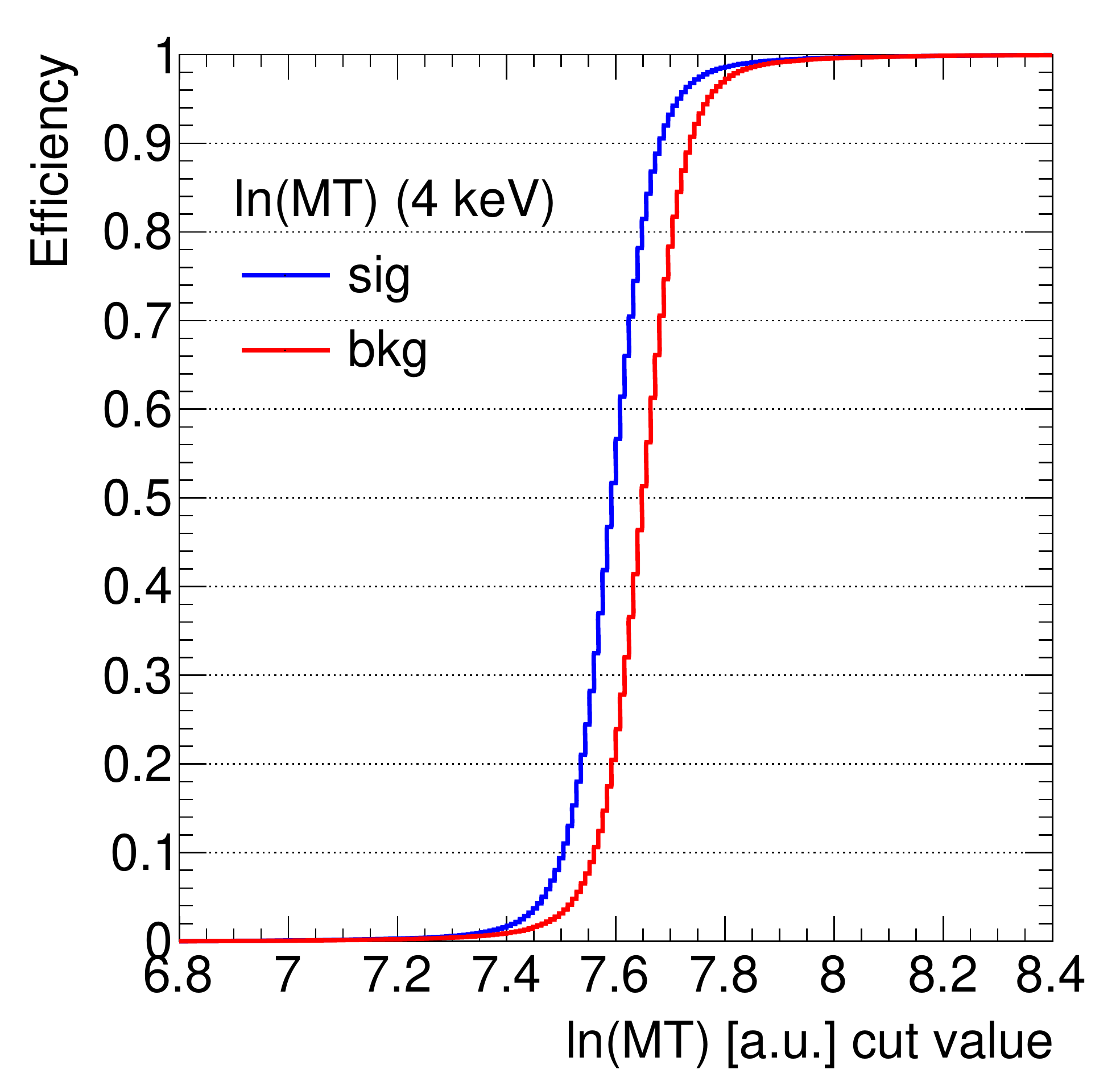}	
		\caption{}
		\label{sigVsbkgEff_lnMT_4keV} 
	\end{subfigure}%
	\begin{subfigure}{0.33333\textwidth}
		\centering
		\includegraphics[height=50mm]{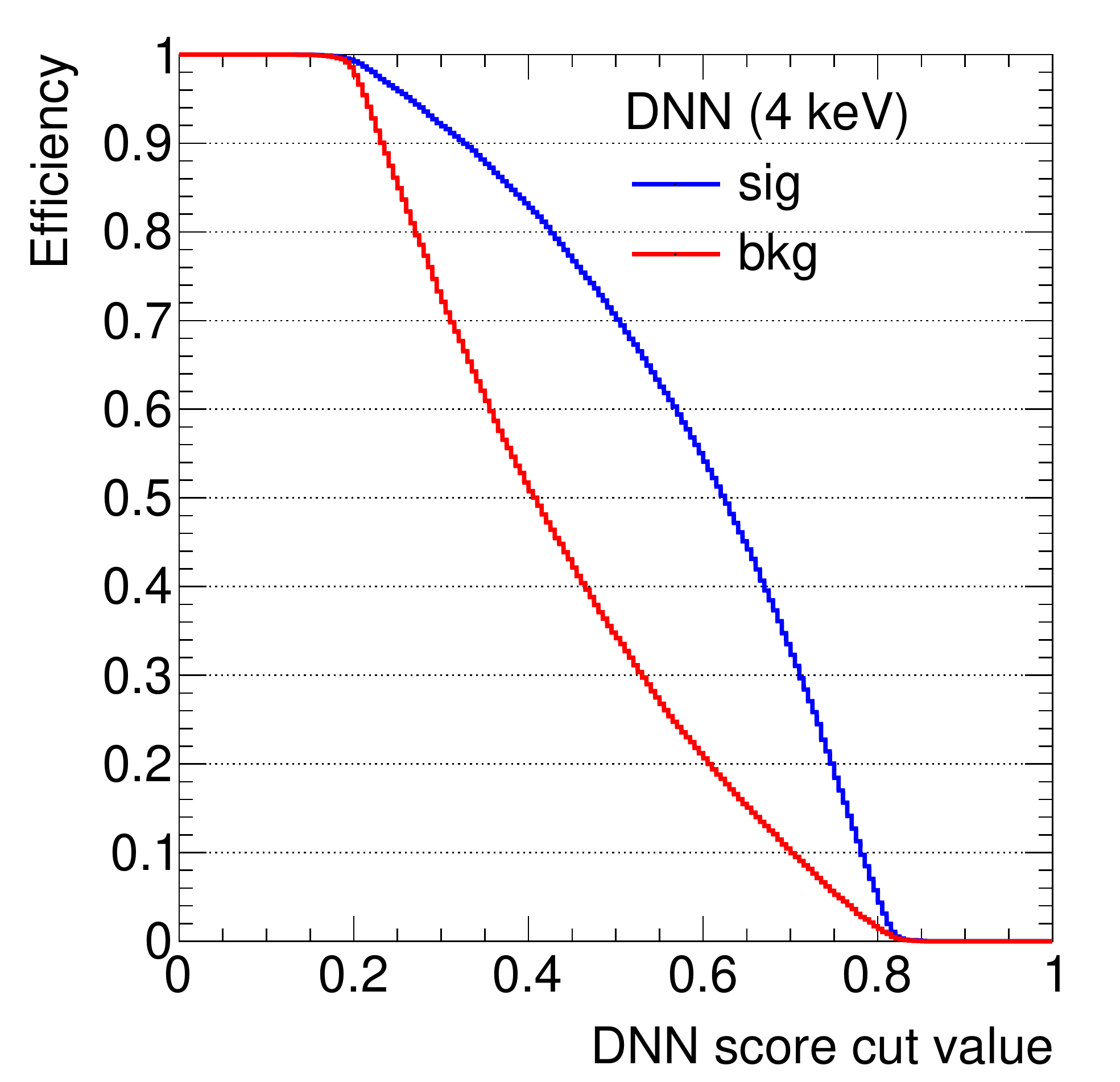}
		\caption{}
		\label{sigVsbkgEff_scoreDNN_4keV} 
	\end{subfigure}%
	\begin{subfigure}{0.33333\textwidth}
		\centering
		\includegraphics[height=50mm]{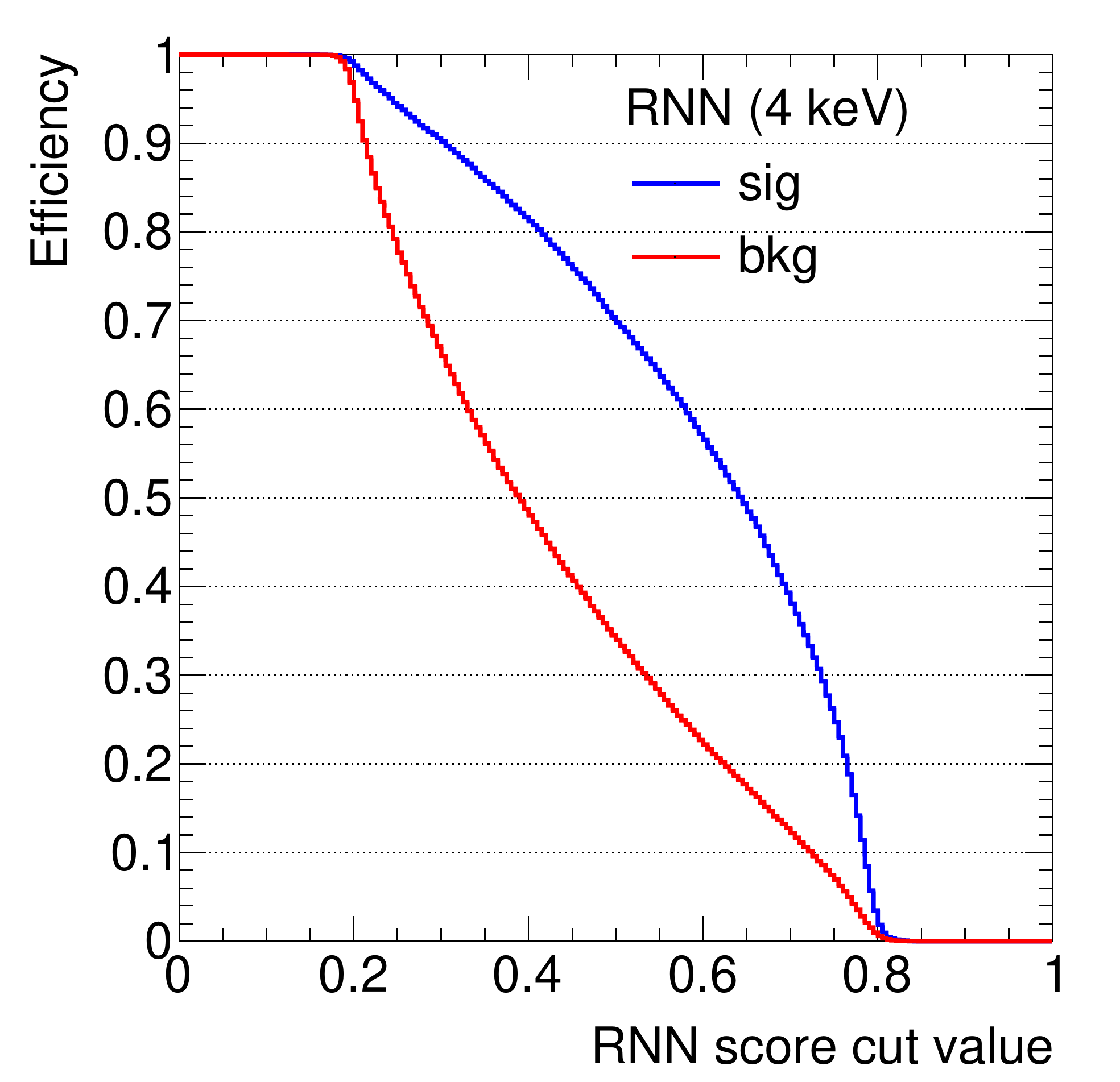}
		\caption{}
		\label{sigVsbkgEff_scoreRNN_4keV} 
	\end{subfigure}
    \caption{Signal and Background efficiency vs. Discriminating Threshold for 4 keV. a) Mean Time. b) DNN. c) RNN.}
	\label{sigVsbkgEff_lnMT_NN_4keV}
\end{figure}

\noindent
The natural log of mean-time, has a good discrimination at low energy deposits \cite{Lee2014}. The expression for the mean time is given by,
\begin{equation}
    ln(MT) = ln\left(\frac{\sum_{i}A_it_i}{\sum_{i}A_i}\right)
\end{equation}
where, $A_i$ is the amplitude at time $t_i$. The distribution of the mean time (after taking the natural log) is shown in figure \ref{lnMT}, while figure \ref{ROC_lnMT_NN} shows the corresponding ROC plot. Although, the difference between the AUC values of the mean time and the network is very small, but if we take a look at the variation of the signal and background efficiencies with the value of the discriminating threshold in figures \ref{sigVsbkgEff_lnMT_NN_20keV} and \ref{sigVsbkgEff_lnMT_NN_4keV}, it can be observed that the signal efficiency remains stable while the background efficiency drops quickly for the networks compared to the mean time. The networks thus, do provide an advantage over mean-time method.\\  

\begin{figure}[htbp!]
    \centering
    \begin{subfigure}{0.5\textwidth}
        \includegraphics[height=58mm]{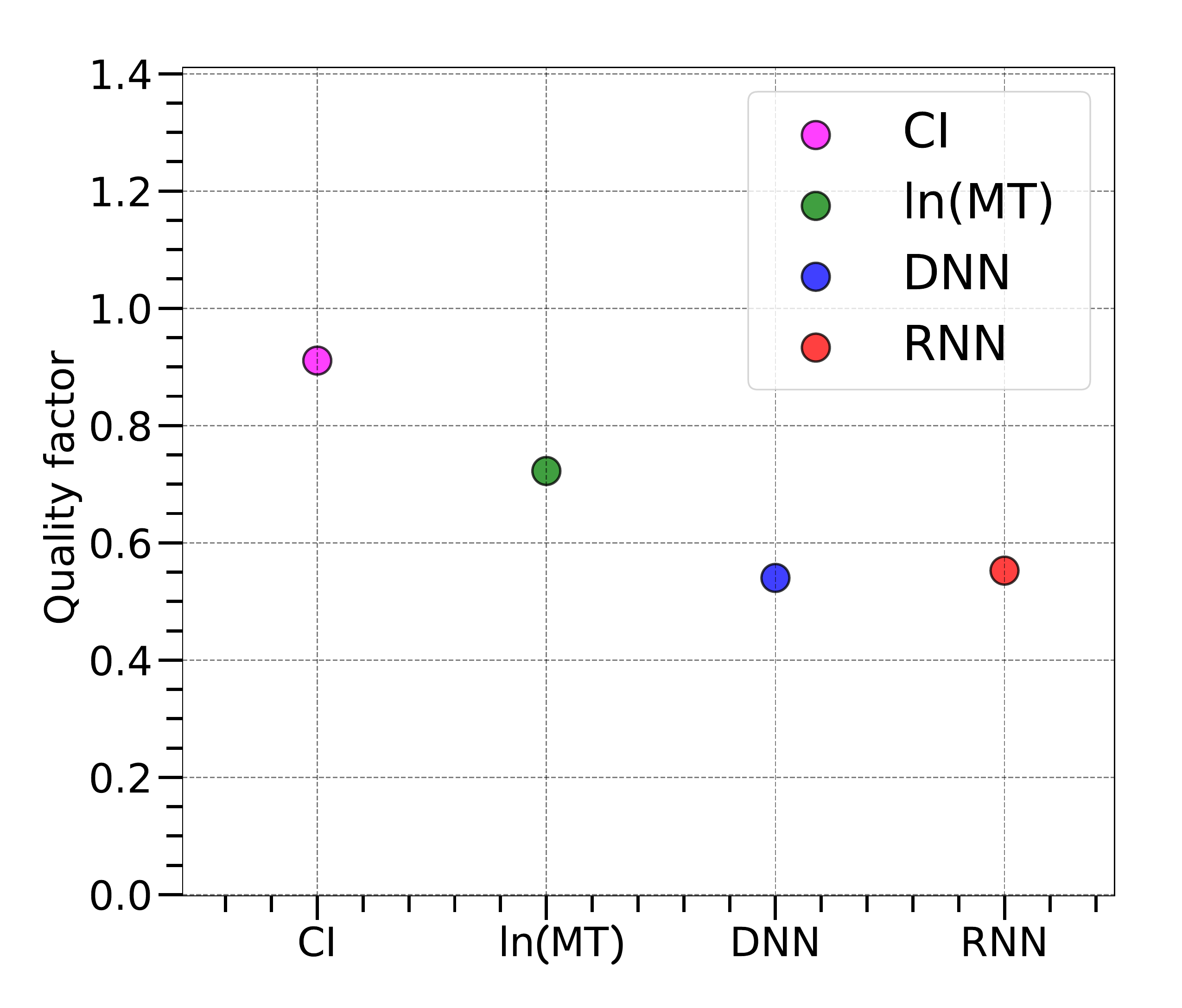}	
        \caption{}
        \label{QF_20keV_0p95}
    \end{subfigure}%
    \begin{subfigure}{0.5\textwidth}
        \includegraphics[height=58mm]{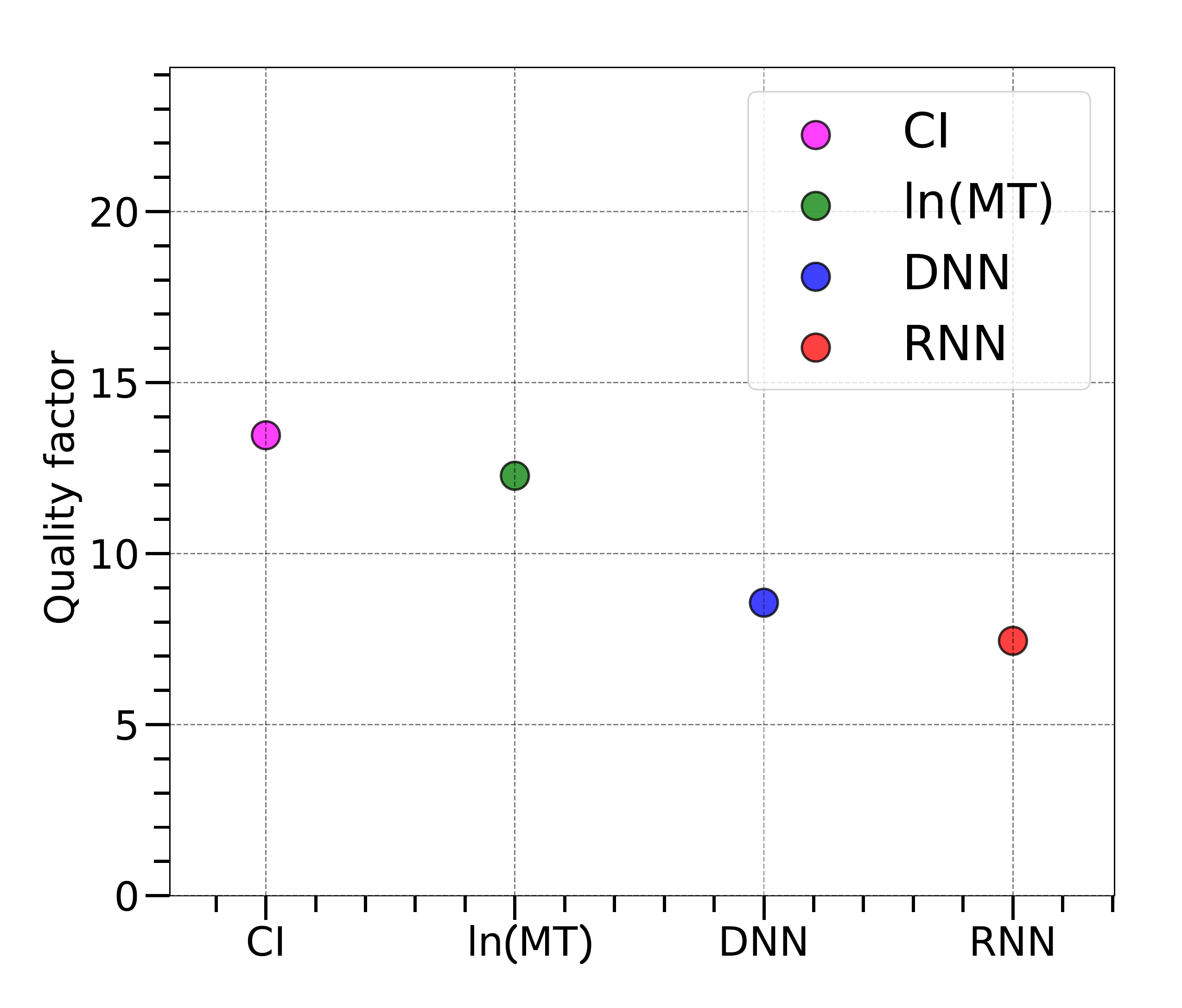}	
        \caption{}
        \label{QF_4keV_0p95}
    \end{subfigure}%
	\caption{Quality factor at 95\% signal efficiency for a) 20 keV and b) 4 keV.}
	\label{qualityFactorPlot} 
\end{figure}

\noindent
The comparison is also done by calculating another figure-of-merit, the quality factor (QF) \cite{Lee2014}, which is given by the relation,
\begin{equation}
    K = \frac{\beta(1-\beta)}{(\alpha-\beta)^{2}}
    \label{qualityFactorEq}
\end{equation}
where, $\alpha$ and $\beta$ are the fraction of signal and background events that satisfy the selection criteria. The best or ideal scenario is when $\alpha = 1$ and $\beta = 0$ which gives the value of $K = 0$. The quality factor is calculated at a signal efficiency of 95\% i.e. $\alpha = 0.95$. Figure \ref{qualityFactorPlot} shows the quality factors calculated for the different methods. The quality factor plots corroborate with the ROC plots that the networks perform better than the conventional methods and also shows that the RNN has an edge over the DNN at low recoil energies. The flip in case of the 20 keV is due to some statistical fluctuation as mentioned earlier.

\section{Conclusion}
\label{conclusion}

\noindent 
We have demonstrated the efficacy of ML-based algorithms for PSD at low recoil energies. The ML methods were compared with the conventional methods - charge integration and mean time. The comparison was done by calculating the AUC values from the ROC plots. The methods were applied to the same datasets so as to have a fair comparison. The datasets were generated from simulation using GEANT4 and a private PMT simulation code. The simulation results were validated by comparing with experimental data and was done for the two inorganic crystals - BGO and CsI(Tl). The match between the simulation and experiment was reasonably well. In case of CsI(Tl) the simulation gave a narrower photo-peak as compared to experiment. This is expected due to the presence of electronic noise in the experimental data. The performance of the PSD methods were benchmarked using only CsI(Tl) datasets.\\ 

\begin{figure}[htbp!]
   	\centering
	\includegraphics[height=70mm]{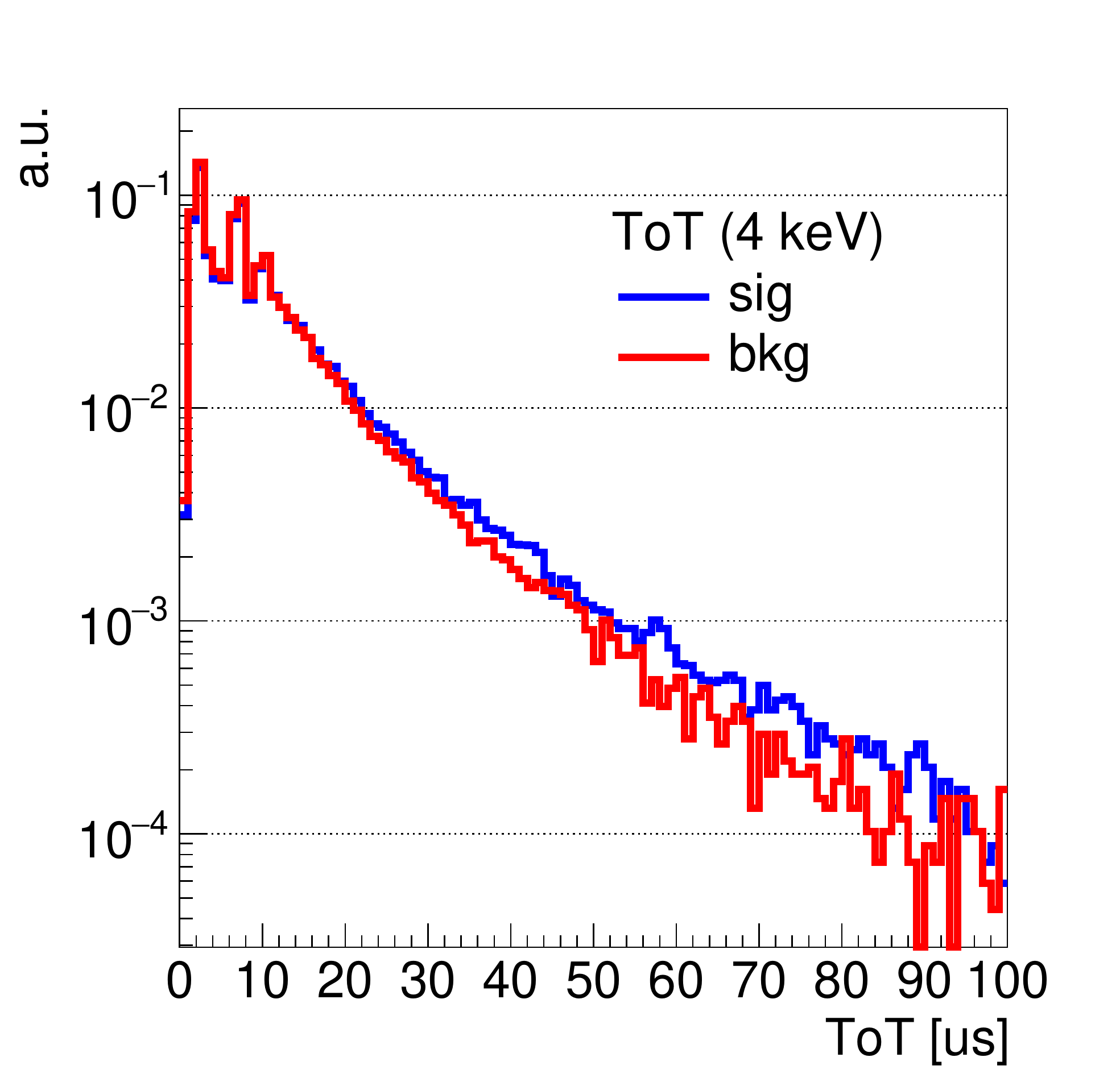}
	\caption{ToT-value distribution for signal and background for 4 keV. Similar distribution is also obtained for 20 keV. }
	\label{ToTvalue}
\end{figure}

\noindent
The ML methods showed an improvement compared to the conventional methods in all the cases. We also tried comparing the results with Time-over-Threshold (ToT) method. But at such low energies the distribution of the signal and background were found to overlap and hence, failed to provide any discrimination. The distributions of the ToT-value for signal and background are shown in figure \ref{ToTvalue}. \\

\noindent
Two network-based ML methods were analyzed in this paper - Dense Neural Network (DNN) and Recurrent Neural Network (RNN). The RNN, which is more adapted for time-series data, did not show marked improvement when compared with DNN. The AUC and quality factor (QF) values are tabulated in Table \ref{resultsSummary} which also includes the conventional methods. The best among the three is taken in case of charge integration (CI) method. \\

\begin{table}[ht!]
\begin{center}
    \begin{tabular}{|c|| c| c| c| c|}
    \hline
    Method & \multicolumn{2}{c|}{AUC} & \multicolumn{2}{c|}{QF} \\
    \hline
    {}   & 20 keV & 4 keV  & 20 keV & 4 keV \\
    \hline
    CI   & 0.0954 & 0.2970 & 0.9108 & 13.46 \\
    lnMT & 0.0776 & 0.2813 & 0.7225 & 12.28 \\   
    DNN  & 0.0683 & 0.2683 & 0.5401 &  8.56 \\       
    RNN  & 0.0696 & 0.2649 & 0.5525 &  7.45 \\ 
    \hline
    \end{tabular}
    \caption{AUC and QF values for the different methods of PSD}
    \label{resultsSummary}
\end{center}
\end{table}

\noindent
It can be concluded that ML methods are capable of performing better at very low recoil energies. The next step would be to apply and test the ML methods in a real-world scenario.

\newpage

\begin{appendices}

\section{Variation of Input with Simulation Parameters}
\label{App_VariationOfInput}

\begin{figure}[htbp!]
    \begin{subfigure}{0.33333\textwidth}
		\centering
		\includegraphics[height=45mm]{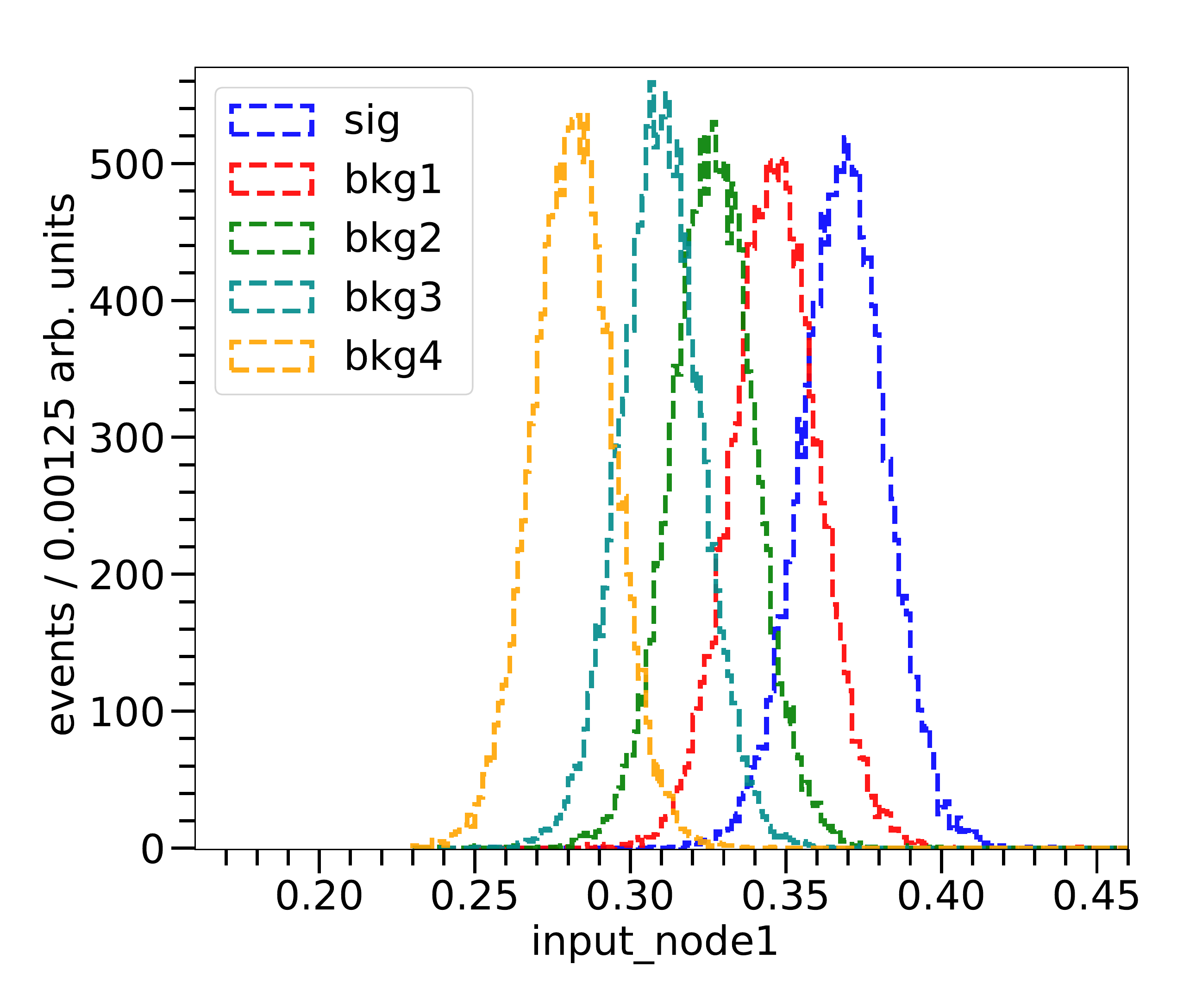}
		\caption{}
		\label{decayTime_node_1} 
	\end{subfigure}%
	\begin{subfigure}{0.33333\textwidth}
		\centering
		\includegraphics[height=45mm]{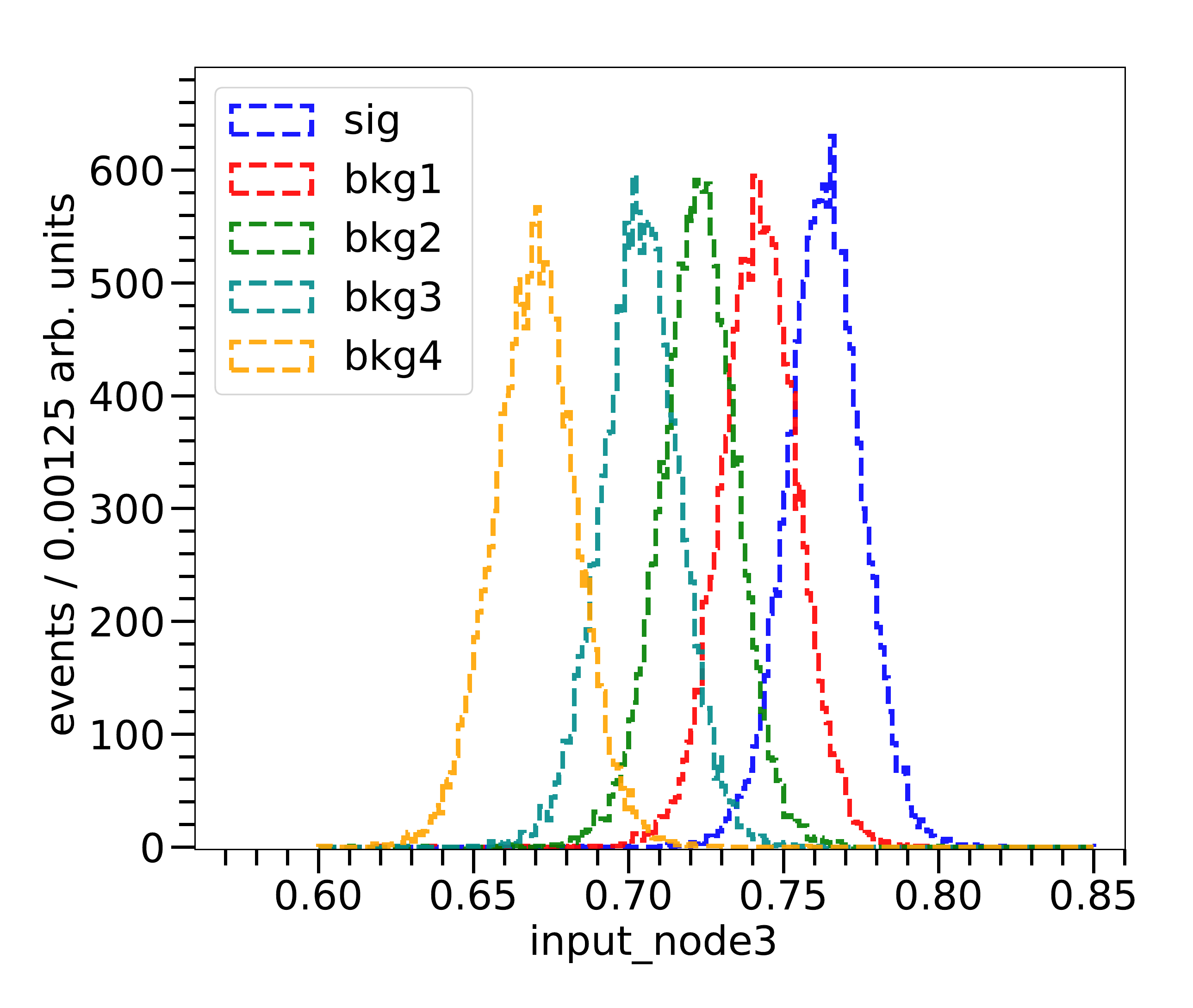}
		\caption{}
		\label{decayTime_node_3} 
	\end{subfigure}%
	\begin{subfigure}{0.33333\textwidth}
		\centering
		\includegraphics[height=45mm]{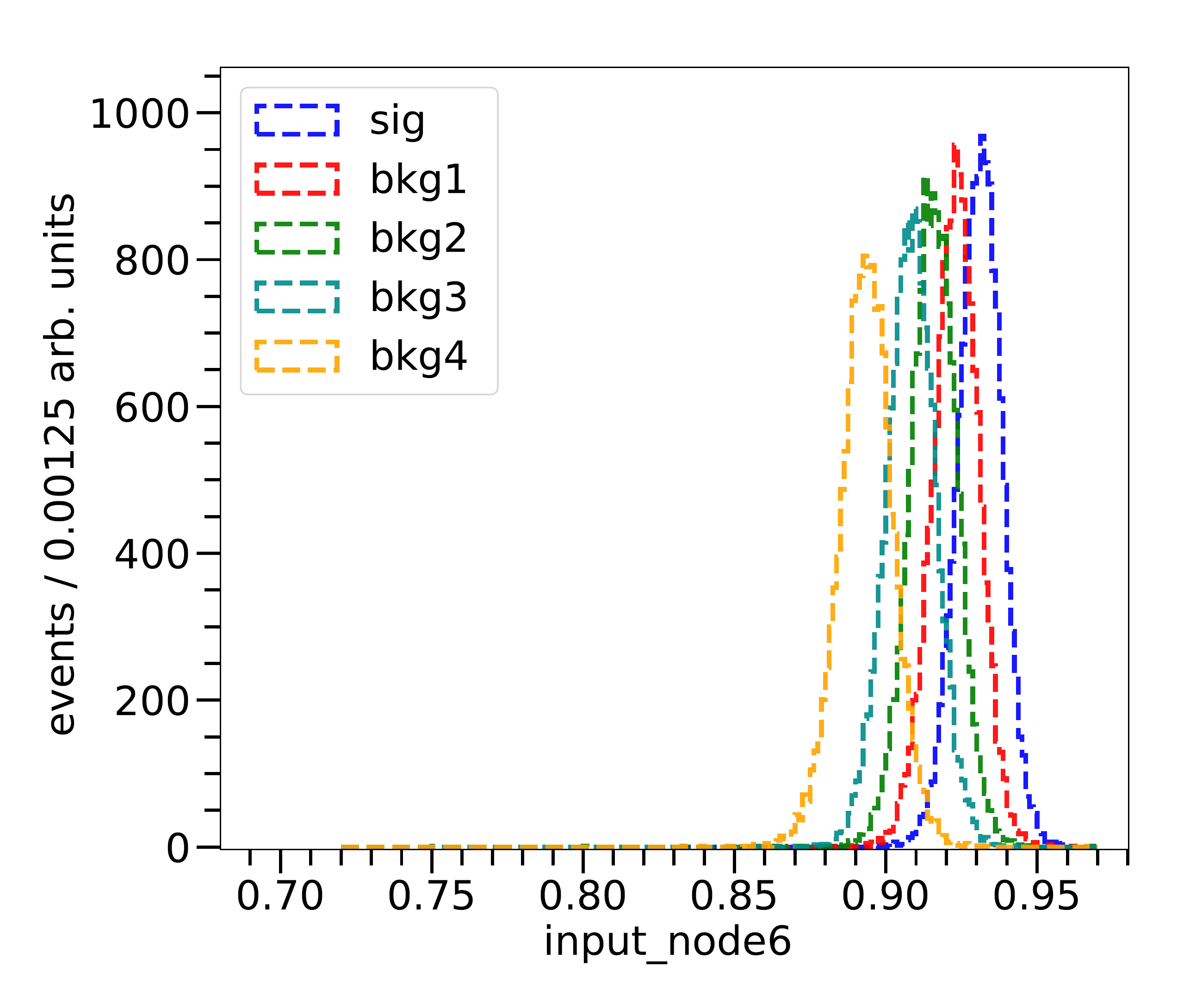}
		\caption{}
		\label{decayTime_node_6} 
	\end{subfigure}%
    \caption{Variation of input distribution as the decay time is increased by 10\% (bkg1), 20\% (bkg2), 30\% (bkg3) and 50\% (bkg4) for a) node 1 b) node 3 and c) node 6. The label `sig' corresponds to 0\% case. Note that in the analysis presented here, `bkg1' has been taken as the background.}
	\label{decayTimeVariation}
\end{figure}

\noindent
The main parameters of the scintillation simulation in GEANT4 are the scintillation spectrum, the rise-time, the fast and slow decay-time constants and the light yield. The scintillation spectrum do not affect the pulse shape, since it only determines the fraction of scintillation photons having a particular wavelength, and doesn't affect the total number of photons generated. The parameters that dominate the pulse shape are the fast and slow decay-time constants as can be seen in figure \ref{decayTimeVariation}. Here, we have plotted the distribution of the input to nodes 1, 3 and 6 that is fed to the DNN for 20 keV recoil energy. The remaining nodes show similar behaviour. In all the following plots, the x-axis is the value (corresponding to an event) seen by the node number mentioned in the x-axis label for each of the cases.\\

\begin{figure}[ht!]
    \begin{subfigure}{0.5\textwidth}
		\centering
		\includegraphics[height=50mm]{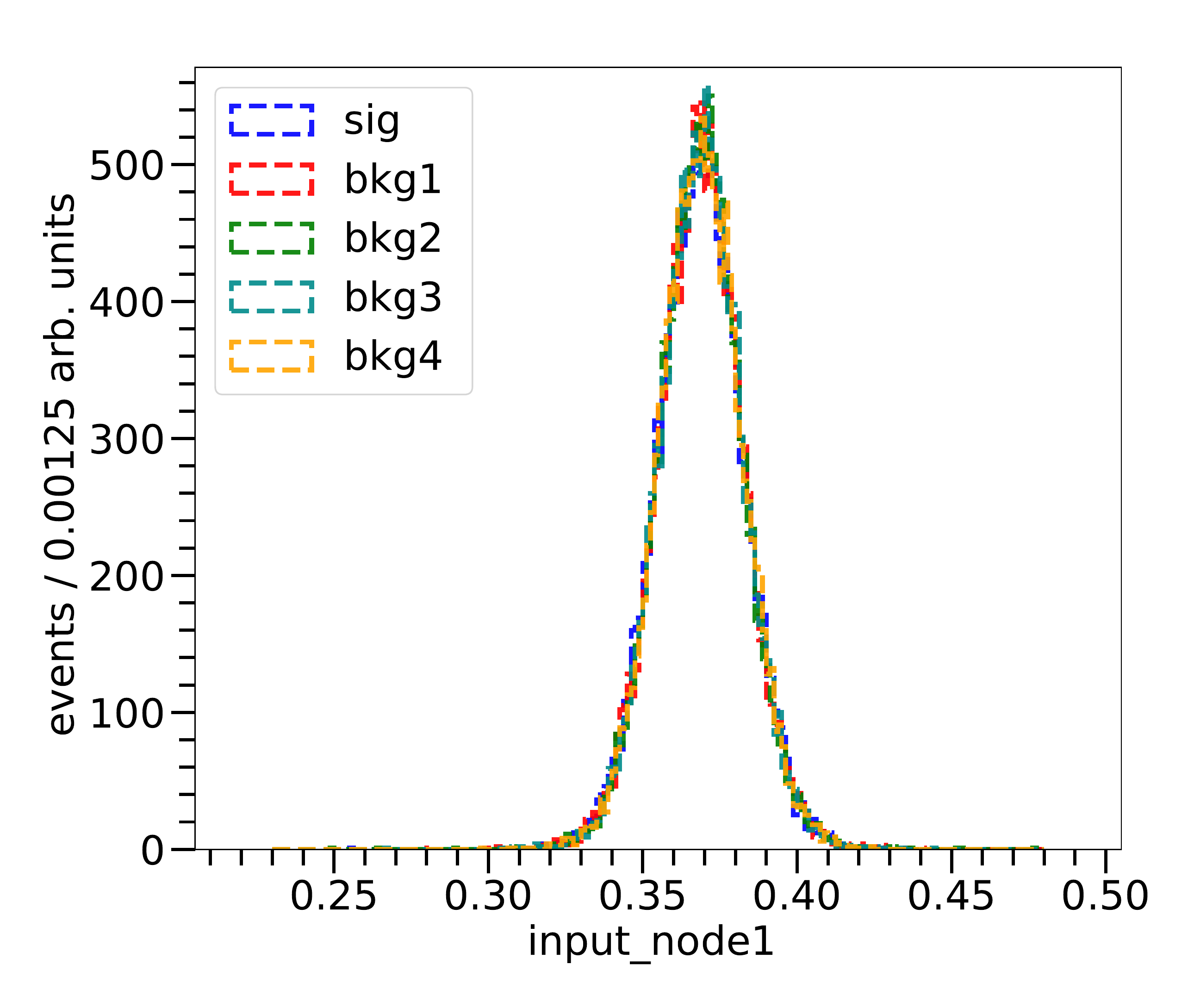}
		\caption{}
		\label{riseTime_node_1} 
	\end{subfigure}%
    \begin{subfigure}{0.5\textwidth}
		\centering
		\includegraphics[height=50mm]{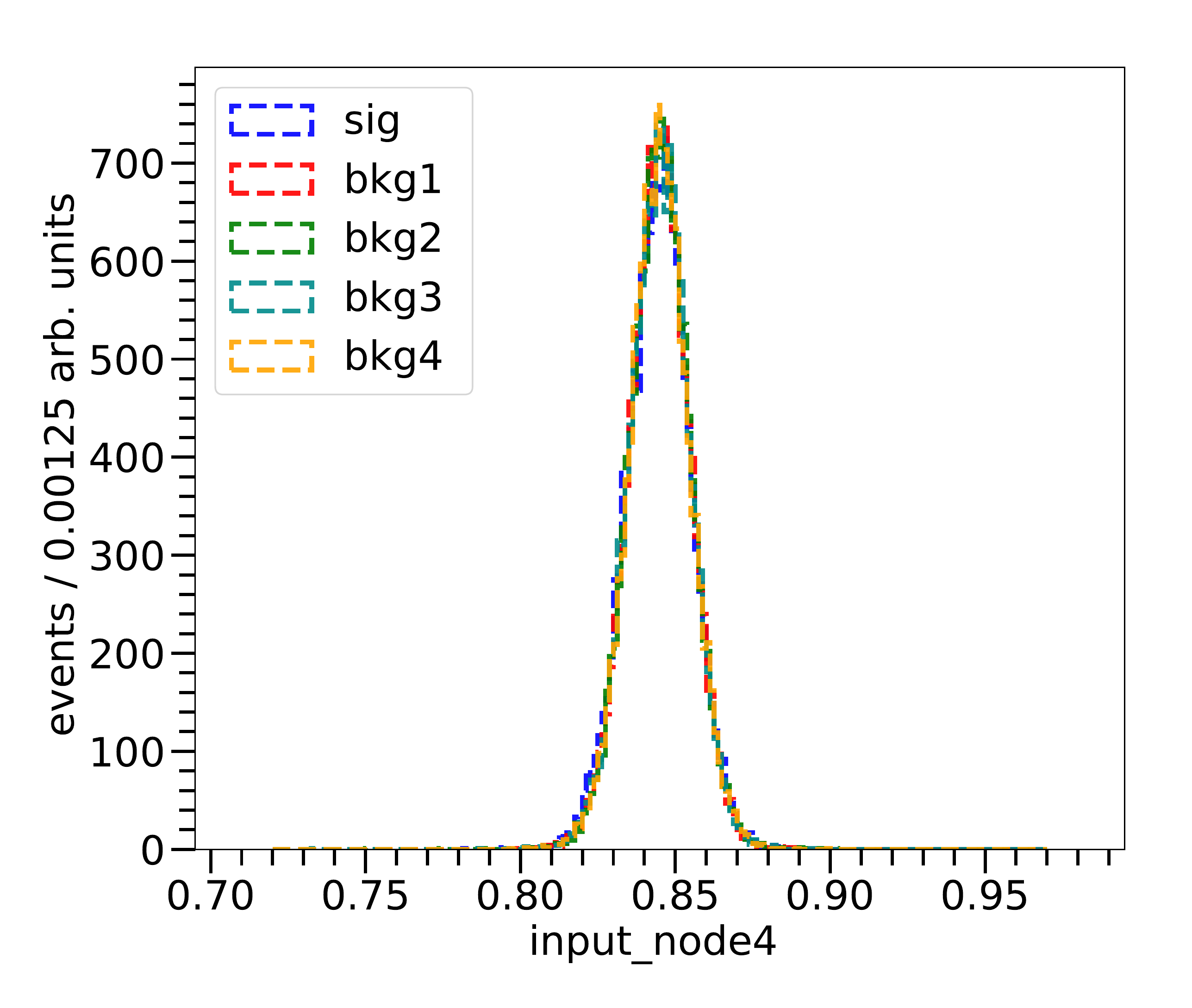}
		\caption{}
		\label{riseTime_node_4} 
	\end{subfigure}%
    \caption{Variation of input distribution as the rise time is increased by 10\% (bkg1), 20\% (bkg2), 30\% (bkg3) and 50\% (bkg4) for a) node 1 b) node 4. All the other nodes show similar behaviour. The label `sig' corresponds to 0\% case.}
	\label{riseTimeVariation}
\end{figure}

\noindent
The rise-time, in principle can affect the pulse shape and hence the input distribution, but the rise time in case of CsI(Tl) is an order less compared to the decay time. Hence, the input distribution varies very little with the rise-time as shown in figure \ref{riseTimeVariation}. \\

\begin{figure}[ht!]
    \begin{subfigure}{0.5\textwidth}
		\centering
		\includegraphics[height=50mm]{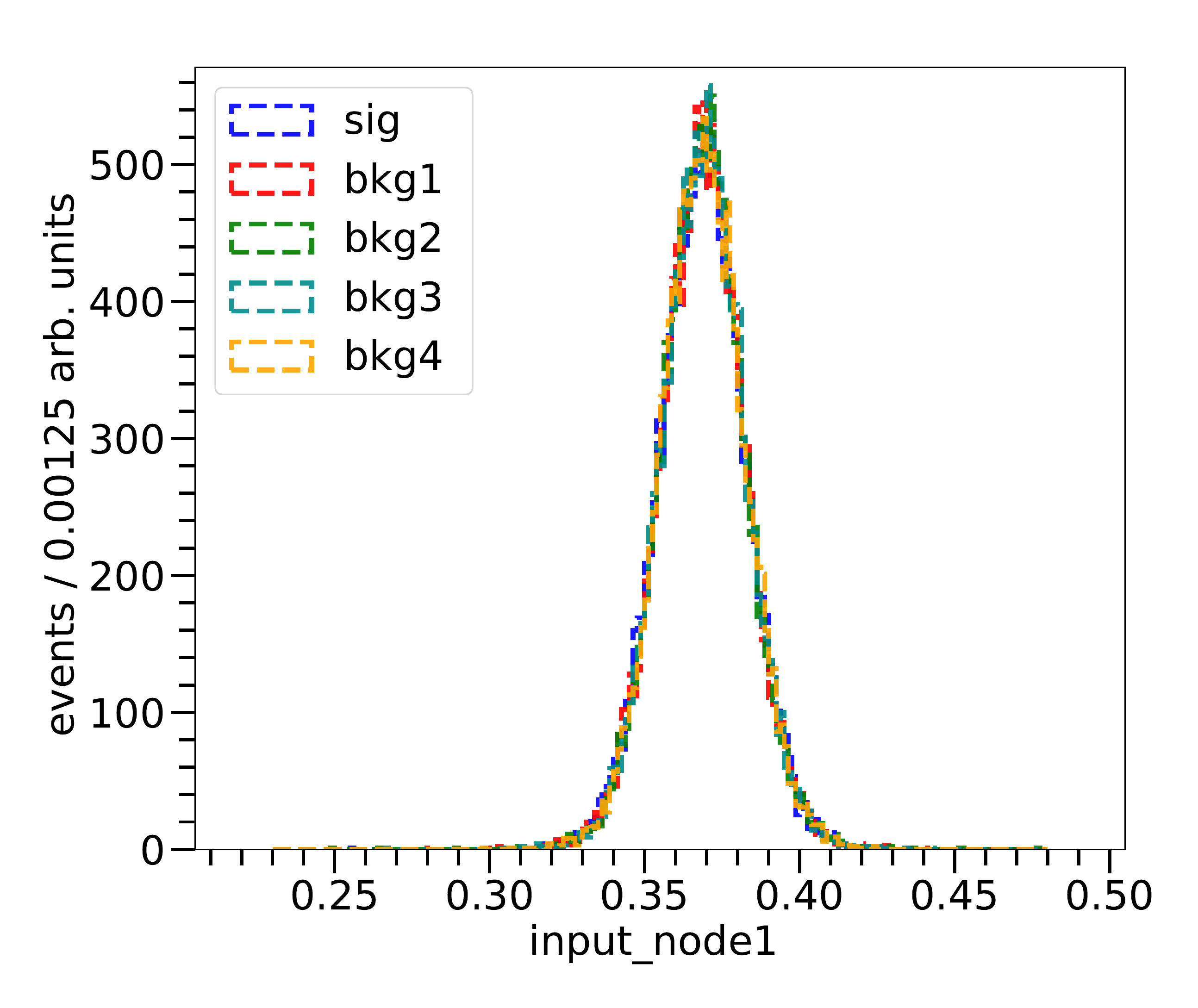}
		\caption{}
		\label{lightYield_node_1} 
	\end{subfigure}%
    \begin{subfigure}{0.5\textwidth}
		\centering
		\includegraphics[height=50mm]{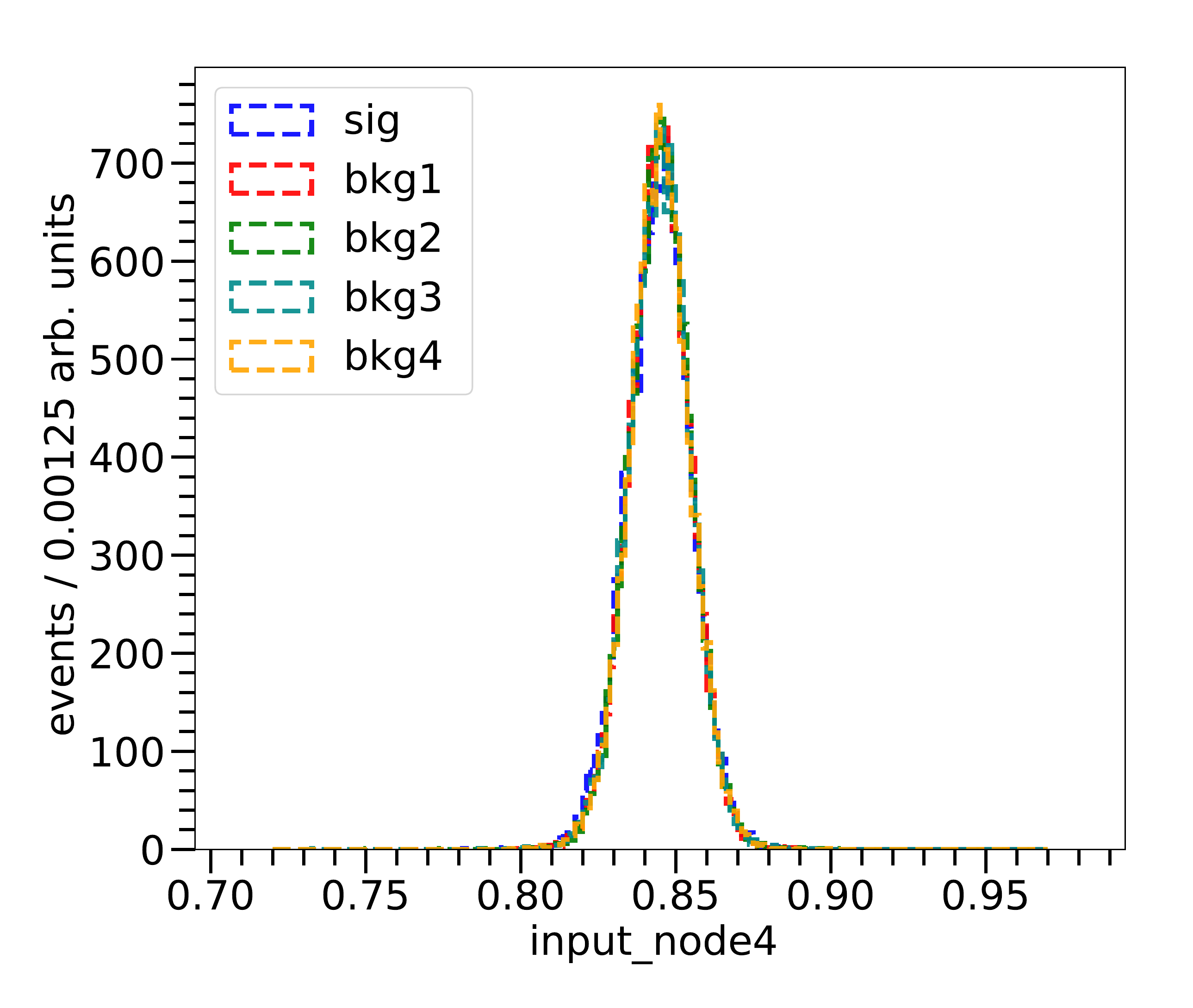}
		\caption{}
		\label{lightYield_node_4} 
	\end{subfigure}%
    \caption{Variation of input distribution as the light yield is increased by 10\% (bkg1), 20\% (bkg2), 30\% (bkg3) and decreased by 10\% (bkg4) for a) node 1 b) node 4. All the other nodes show similar behaviour. The label `sig' corresponds to 0\% case.}
	\label{lightYieldVariation}
\end{figure}

\noindent
The light yield governs the number of scintillation photons generated per unit keV of energy deposited in the scintillator by the incident particle. As such, it only increases the pulse height, keeping the pulse profile unchanged. Figure \ref{lightYieldVariation} shows the effect of varying the light yield parameter on the input distribution.

\newpage
\phantom{skip}
\newpage
\section{Loss Curves}
\label{App_lossCurves}

\begin{figure}[ht!]
    \begin{subfigure}{0.5\textwidth}
		\centering
		\includegraphics[height=60mm]{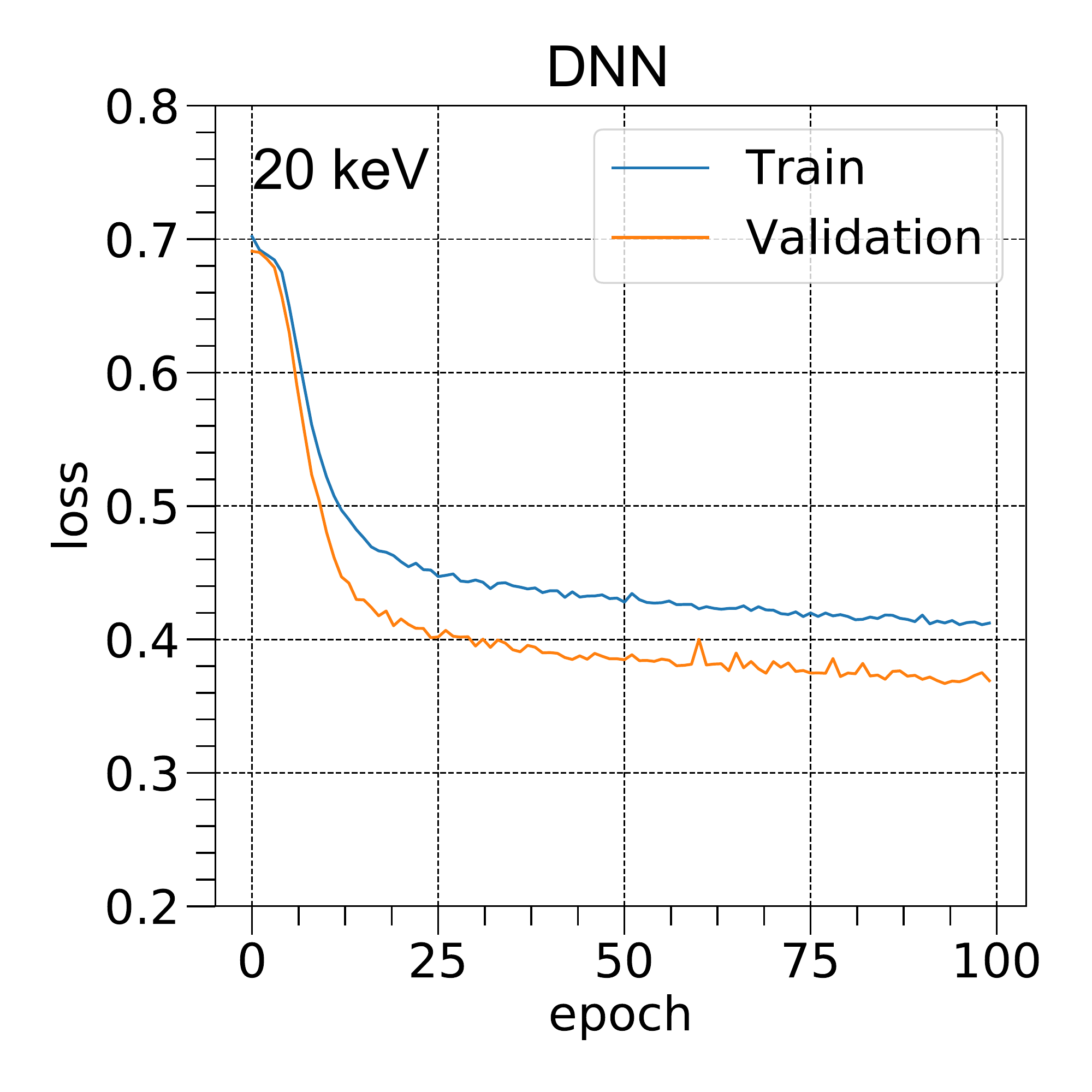}
		\caption{}
		\label{DNNmodelLoss_20keV} 
	\end{subfigure}%
    \begin{subfigure}{0.5\textwidth}
		\centering
		\includegraphics[height=60mm]{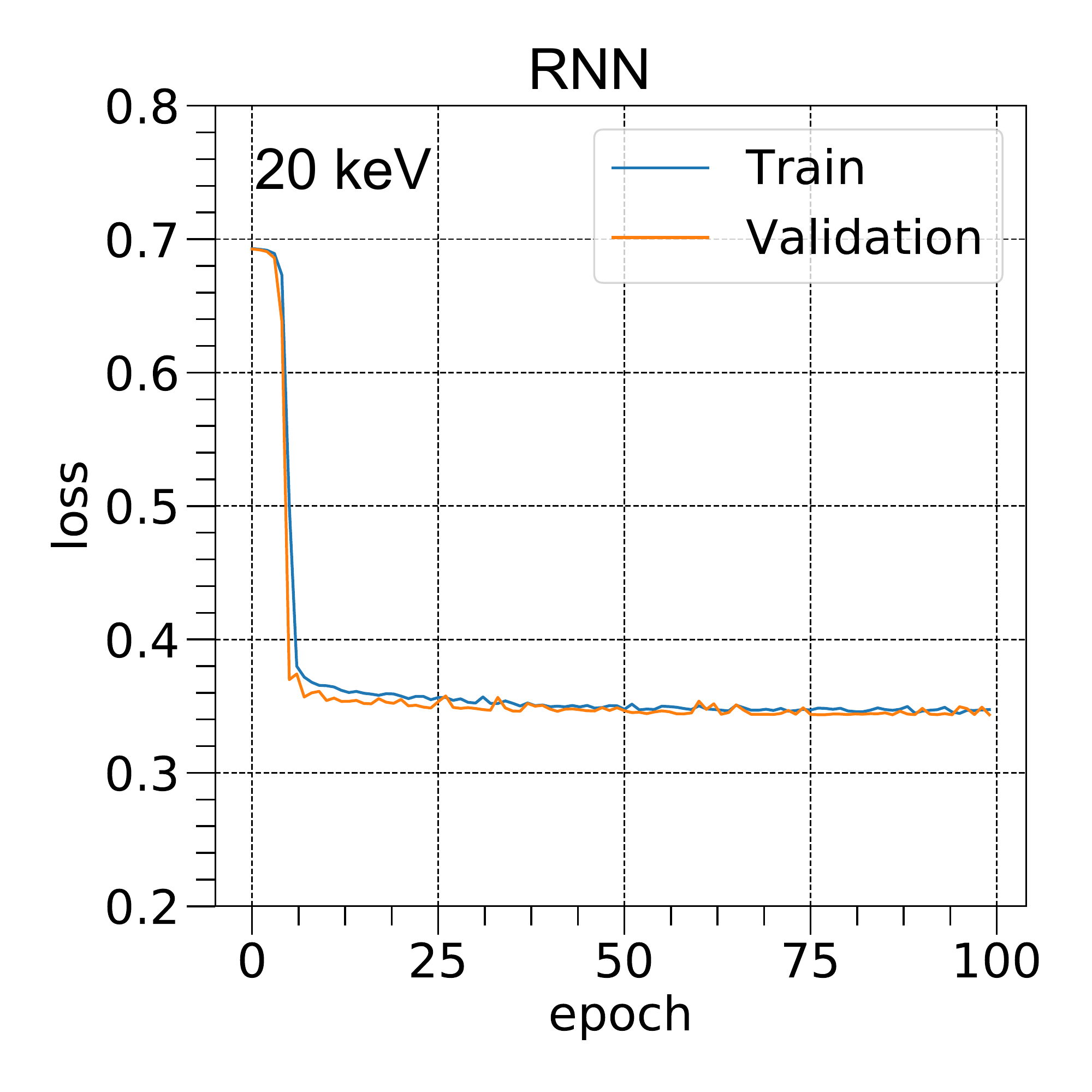}
		\caption{}
		\label{RNNmodelLoss_20keV} 
	\end{subfigure}%
    \caption{Evolution of the loss function with training epoch for 20 keV recoil energy}
	\label{lossCurves_20keV}
\end{figure}

\begin{figure}[ht!]
    \begin{subfigure}{0.5\textwidth}
		\centering
		\includegraphics[height=60mm]{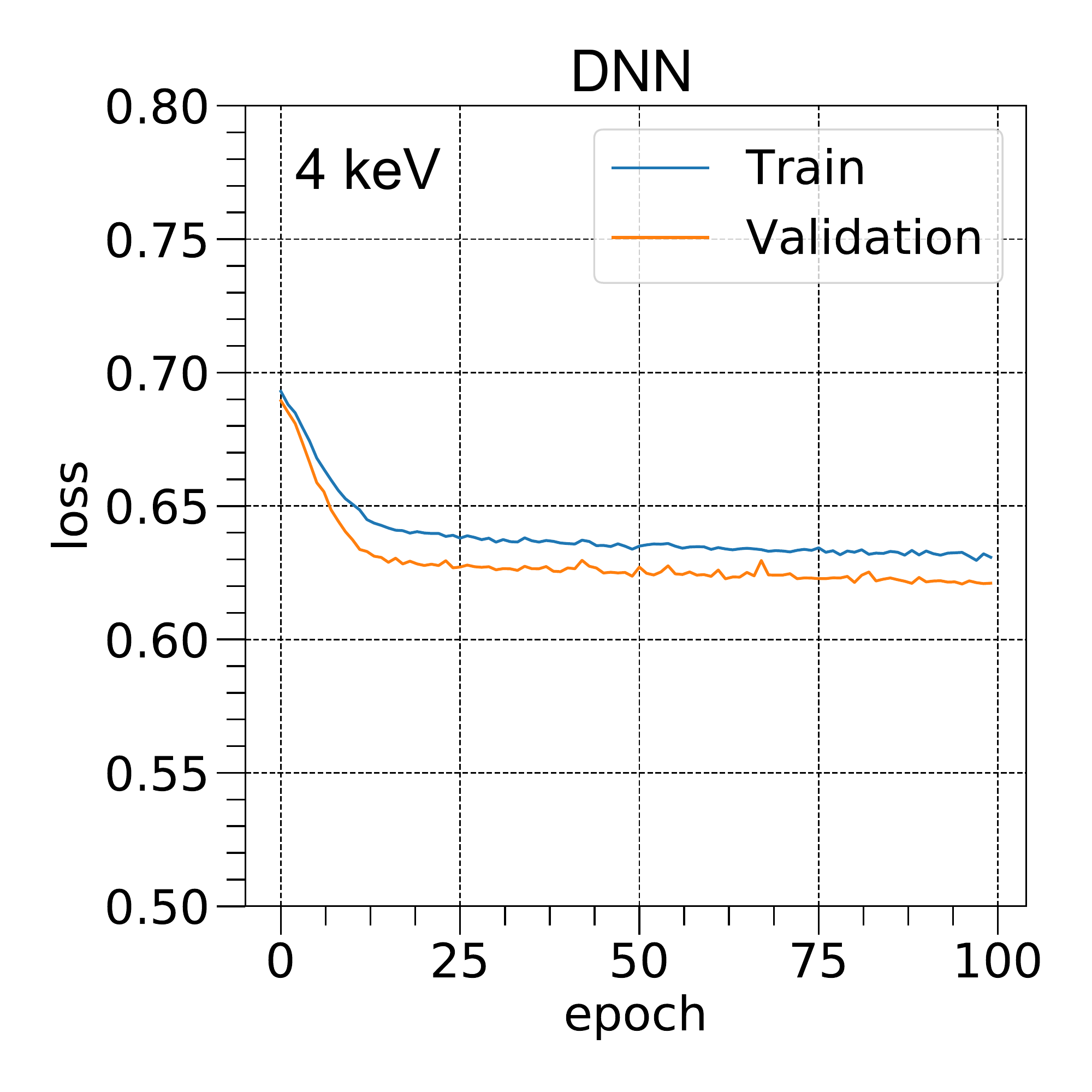}
		\caption{}
		\label{DNNmodelLoss_4keV} 
	\end{subfigure}%
    \begin{subfigure}{0.5\textwidth}
		\centering
		\includegraphics[height=60mm]{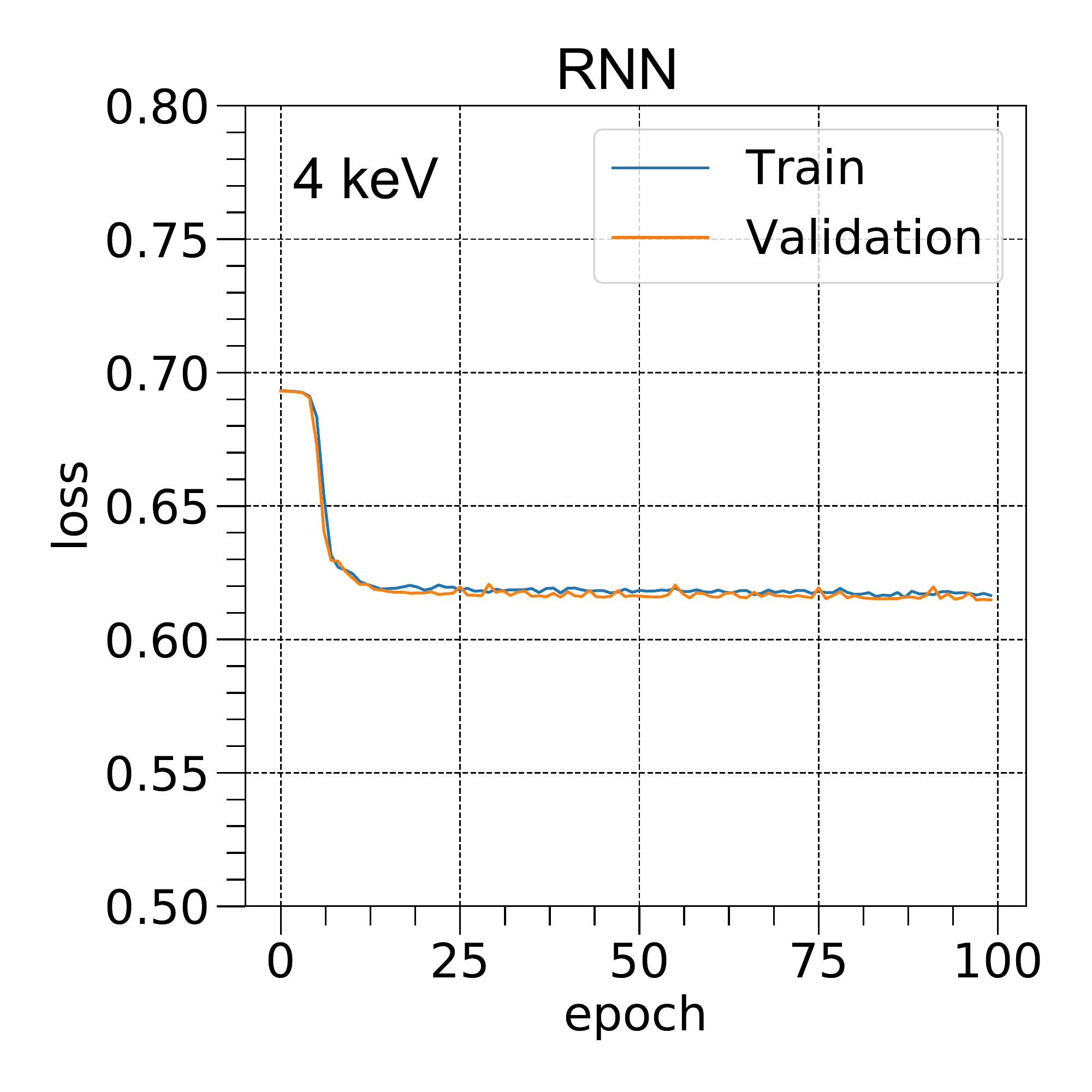}
		\caption{}
		\label{RNNmodelLoss_4keV} 
	\end{subfigure}%
    \caption{Evolution of the loss function with training epoch for 4 keV recoil energy}
	\label{lossCurves_4keV}
\end{figure}

Figures \ref{lossCurves_20keV} and \ref{lossCurves_4keV} show how the loss function changes with each training epoch for 20 keV and 4 keV recoil energy respectively. In both the cases it can be observed that the loss function saturates around the same value, but if looked upon closely then the RNN has slightly lower value after 100 epochs. Thus, the DNN is able to learn the  features of the dataset almost at par with the RNN. However, the change in loss function is very rapid in case of RNN compared to DNN which reckons that RNN is more optimized in learning the features of the time-series dataset.

\end{appendices}

\newpage
\bibliography{references}

\providecommand{\href}[2]{#2}\begingroup\raggedright\begin{thebibliography}{10}

\bibitem{Storey1958}
R.S.~Storey, W.~Jack and A.~Ward, \emph{The fluorescent decay of {CsI}(tl) for
  particles of different ionization density},
  \href{https://doi.org/10.1088/0370-1328/72/1/302}{\emph{Proceedings of the
  Physical Society} {\bfseries 72} (1958) 1}.

\bibitem{CImethod_Sabbah1968}
B.~Sabbah and A.~Suhami, \emph{An accurate pulse-shape discriminator for a wide
  range of energies},
  \href{https://doi.org/https://doi.org/10.1016/0029-554X(68)90035-9}{\emph{Nuclear
  Instruments and Methods} {\bfseries 58} (1968) 102}.

\bibitem{Lee2014}
H.S.~Lee et~al., \emph{{Neutron calibration facility with an Am-Be source for
  pulse shape discrimination measurement of CsI(Tl) crystals}},
  \href{https://doi.org/10.1088/1748-0221/9/11/P11015}{\emph{JINST} {\bfseries
  9} (2014) P11015} [\href{https://arxiv.org/abs/1409.0948}{{\ttfamily
  1409.0948}}].

\bibitem{ZeroX_Roush1964}
M.~Roush, M.~Wilson and W.~Hornyak, \emph{Pulse shape discrimination},
  \href{https://doi.org/https://doi.org/10.1016/0029-554X(64)90333-7}{\emph{Nuclear
  Instruments and Methods} {\bfseries 31} (1964) 112}.

\bibitem{PGA_DMellow2007}
B.~D'Mellow, M.~Aspinall, R.~Mackin, M.~Joyce and A.~Peyton, \emph{Digital
  discrimination of neutrons and gamma-rays in liquid scintillators using pulse
  gradient analysis},
  \href{https://doi.org/https://doi.org/10.1016/j.nima.2007.04.174}{\emph{Nuclear
  Instruments and Methods in Physics Research Section A: Accelerators,
  Spectrometers, Detectors and Associated Equipment} {\bfseries 578} (2007)
  191}.

\bibitem{ToT_Kipnis1997}
I.~Kipnis, T.~Collins, J.~DeWitt, S.~Dow, A.~Frey, A.~Grillo et~al., \emph{A
  time-over-threshold machine: the readout integrated circuit for the babar
  silicon vertex tracker}, \href{https://doi.org/10.1109/23.603658}{\emph{IEEE
  Transactions on Nuclear Science} {\bfseries 44} (1997) 289}.

\bibitem{ToT_Ngyren1991}
D.~Nygren, \emph{Converting vice to virtue: can time-walk be used as a measure
  of deposited charge in silicon detectors?}, {\emph{Internal LBL note} (1991)
  }.

\bibitem{Yousefi2009}
S.~Yousefi, L.~Lucchese and M.~Aspinall, \emph{Digital discrimination of
  neutrons and gamma-rays in liquid scintillators using wavelets},
  \href{https://doi.org/https://doi.org/10.1016/j.nima.2008.09.028}{\emph{Nuclear
  Instruments and Methods in Physics Research Section A: Accelerators,
  Spectrometers, Detectors and Associated Equipment} {\bfseries 598} (2009)
  551}.

\bibitem{CHIMERA2002}
M.~Alderighi, A.~Anzalone, R.~Basssini, I.~Berceanu, J.~Blicharska, C.~Boiano
  et~al., \emph{Particle identification method in the csi(tl) scintillator used
  for the chimera 4$\pi$ detector},
  \href{https://doi.org/https://doi.org/10.1016/S0168-9002(02)00800-8}{\emph{Nuclear
  Instruments and Methods in Physics Research Section A: Accelerators,
  Spectrometers, Detectors and Associated Equipment} {\bfseries 489} (2002)
  257}.

\bibitem{Longo2020}
S.~Longo, J.~Roney, C.~Cecchi, S.~Cunliffe, T.~Ferber, H.~Hayashii et~al.,
  \emph{Csi(tl) pulse shape discrimination with the belle ii electromagnetic
  calorimeter as a novel method to improve particle identification at
  electron–positron colliders},
  \href{https://doi.org/https://doi.org/10.1016/j.nima.2020.164562}{\emph{Nuclear
  Instruments and Methods in Physics Research Section A: Accelerators,
  Spectrometers, Detectors and Associated Equipment} {\bfseries 982} (2020)
  164562}.

\bibitem{KIMS2002}
H.~Park, D.~Choi, J.~Choi, I.~Hahn, M.~Hwang, W.~Kang et~al., \emph{Neutron
  beam test of csi crystal for dark matter search},
  \href{https://doi.org/https://doi.org/10.1016/S0168-9002(02)01274-3}{\emph{Nuclear
  Instruments and Methods in Physics Research Section A: Accelerators,
  Spectrometers, Detectors and Associated Equipment} {\bfseries 491} (2002)
  460}.

\bibitem{Umehara2015}
S.~Umehara, T.~Kishimoto, M.~Nomachi, S.~Ajimura, T.~Iida, K.~Nakajima et~al.,
  \emph{Search for neutrino-less double beta decay with candles},
  \href{https://doi.org/10.1016/j.phpro.2014.12.046}{\emph{Physics Procedia}
  {\bfseries 61} (2015) 283}.

\bibitem{Ghosh2022}
S.~Ghosh, S.~Dutta, N.K.~Mondal and S.~Saha, \emph{Measurements of gamma ray,
  cosmic muon and residual neutron background fluxes for rare event search
  experiments at an underground laboratory},
  \href{https://doi.org/https://doi.org/10.1016/j.astropartphys.2022.102700}{\emph{Astroparticle
  Physics} {\bfseries 139} (2022) 102700}.

\bibitem{KIMS2014}
H.~Lee, H.~Bhang, J.~Choi, S.~Choi, I.~Hahn, E.~Jeon et~al., \emph{Neutron
  calibration facility with an am-be source for pulse shape discrimination
  measurement of {CsI}(tl) crystals},
  \href{https://doi.org/10.1088/1748-0221/9/11/p11015}{\emph{Journal of
  Instrumentation} {\bfseries 9} (2014) P11015}.

\bibitem{Freund1997}
Y.~Freund and R.E.~Schapire, \emph{A decision-theoretic generalization of
  on-line learning and an application to boosting},
  \href{https://doi.org/https://doi.org/10.1006/jcss.1997.1504}{\emph{Journal
  of Computer and System Sciences} {\bfseries 55} (1997) 119}.

\bibitem{Goodfellow2016}
I.~Goodfellow, Y.~Bengio and A.~Courville, \emph{Deep Learning}, MIT Press
  (2016).

\bibitem{ImageNet_CNN}
A.~Krizhevsky, I.~Sutskever and G.E.~Hinton, \emph{Imagenet classification with
  deep convolutional neural networks},  in \emph{Advances in Neural Information
  Processing Systems}, F.~Pereira, C.J.C.~Burges, L.~Bottou and
  K.Q.~Weinberger, eds., vol.~25, Curran Associates, Inc., 2012,
  \href{https://proceedings.neurips.cc/paper/2012/file/c399862d3b9d6b76c8436e924a68c45b-Paper.pdf}{https://proceedings.neurips.cc/paper/2012/file/c399862d3b9d6b76c8436e924a68c45b-Paper.pdf}.

\bibitem{CMS2018}
A.D.~Florio, F.~Pantaleo and A.C.~and, \emph{Convolutional neural network for
  track seed filtering at the {CMS} high-level trigger},
  \href{https://doi.org/10.1088/1742-6596/1085/4/042040}{\emph{Journal of
  Physics: Conference Series} {\bfseries 1085} (2018) 042040}.

\bibitem{CMS2020}
{\scshape The CMS} collaboration, \emph{A deep neural network to search for new
  long-lived particles decaying to jets},
  \href{https://doi.org/10.1088/2632-2153/ab9023}{\emph{Machine Learning:
  Science and Technology} {\bfseries 1} (2020) 035012}.

\bibitem{DUNE2020}
{\scshape DUNE} collaboration, \emph{Neutrino interaction classification with a
  convolutional neural network in the dune far detector},
  \href{https://doi.org/10.1103/PhysRevD.102.092003}{\emph{Phys. Rev. D}
  {\bfseries 102} (2020) 092003}.

\bibitem{Aurisano2016}
A.~Aurisano, A.~Radovic, D.~Rocco, A.~Himmel, M.~Messier, E.~Niner et~al.,
  \emph{A convolutional neural network neutrino event classifier},
  \href{https://doi.org/10.1088/1748-0221/11/09/p09001}{\emph{Journal of
  Instrumentation} {\bfseries 11} (2016) P09001}.

\bibitem{Soham2022}
S.~Bhattacharya, M.~Guchait and A.H.~Vijay, \emph{Boosted top quark tagging and
  polarization measurement using machine learning},
  \href{https://doi.org/10.1103/PhysRevD.105.042005}{\emph{Phys. Rev. D}
  {\bfseries 105} (2022) 042005}.

\bibitem{WZTagNN_Chen2020}
Y.-C.J.~Chen, C.-W.~Chiang, G.~Cottin and D.~Shih, \emph{Boosted $w$ and $z$
  tagging with jet charge and deep learning},
  \href{https://doi.org/10.1103/PhysRevD.101.053001}{\emph{Phys. Rev. D}
  {\bfseries 101} (2020) 053001}.

\bibitem{dropout_Srivastava2014}
N.~Srivastava, G.~Hinton, A.~Krizhevsky, I.~Sutskever and R.~Salakhutdinov,
  \emph{Dropout: A simple way to prevent neural networks from overfitting},
  {\emph{Journal of Machine Learning Research} {\bfseries 15} (2014) 1929}.

\bibitem{Nair2010}
V.~Nair and G.E.~Hinton, \emph{Rectified linear units improve restricted
  boltzmann machines},  in \emph{ICML}, 2010.

\bibitem{geant4}
S.~Agostinelli, J.~Allison, K.~Amako, J.~Apostolakis, H.~Araujo, P.~Arce
  et~al., \emph{Geant4—a simulation toolkit},
  \href{https://doi.org/https://doi.org/10.1016/S0168-9002(03)01368-8}{\emph{Nuclear
  Instruments and Methods in Physics Research Section A: Accelerators,
  Spectrometers, Detectors and Associated Equipment} {\bfseries 506} (2003)
  250}.

\bibitem{unifiedModel_ref}
A.~Bismark, \emph{Simulation and characterization of a LXe TPC for DARWIN
  R\&D}, Ph.D. thesis, University of Zurich, 12, 2019.
\newblock 10.6094/UNIFR/154696.

\bibitem{Silva2010}
C.~Silva, J.P.~da~Cunha, A.~Pereira, V.~Chepel, M.I.~Lopes, V.~Solovov et~al.,
  \emph{Reflectance of polytetrafluoroethylene for xenon scintillation light},
  \href{https://doi.org/10.1063/1.3318681}{\emph{Journal of Applied Physics}
  {\bfseries 107} (2010) 064902}.

\bibitem{pmtGain_Rademacker2002}
J.~Rademacker, \emph{{An exact formula to describe the amplification process in
  a photomultiplier tube}},
  \href{https://doi.org/10.1016/S0168-9002(01)02055-1}{\emph{Nucl. Instrum.
  Meth. A} {\bfseries 484} (2002) 432}
  [\href{https://arxiv.org/abs/physics/0406036}{{\ttfamily physics/0406036}}].

\bibitem{Ogawara2016}
R.~Ogawara and M.~Ishikawa, \emph{{Signal pulse emulation for scintillation
  detectors using Geant4 Monte Carlo with light tracking simulation}},
  \href{https://doi.org/10.1063/1.4959186}{\emph{Rev. Sci. Instrum.} {\bfseries
  87} (2016) 075114}.

\end{thebibliography}\endgroup

\end{document}